\documentclass[journal]{IEEEtran}
\IEEEoverridecommandlockouts
\usepackage[utf8]{inputenc}
\usepackage{mathcomSTEv4}
\usepackage{amsmath,amssymb,amsthm}
\usepackage{accents,latexsym,cancel}
\usepackage{cite}

\usepackage[dvipsnames]{xcolor}
\definecolor{myPink}{RGB}{255,105,183}

\usepackage[T1]{fontenc}
\usepackage{graphics} % for pdf, bitmapped graphics files
\usepackage{epsfig} % for postscript graphics files
\usepackage[mathscr]{euscript}
\usepackage{algorithm}
\usepackage[noend]{algpseudocode}
\usepackage{bbm}
\makeatletter
\def\BState{\State\hskip-\ALG@thistlm}
\makeatother

\usepackage{tikz}
\usetikzlibrary{arrows,shapes,chains,matrix,positioning,scopes,patterns,calc,
decorations.markings,
decorations.pathmorphing,
}

\usepackage{pgfplots}
\pgfplotsset{compat=1.3}
\usepgflibrary{shapes}

\newtheorem{theorem}{Theorem}
\newtheorem{lemma}[theorem]{Lemma}
\newtheorem{proposition}[theorem]{Proposition}

\newtheorem{remark}[theorem]{Remark}

\renewcommand{\epsilon}{\varepsilon}

\newcommand{\RNum}[1]{\uppercase\expandafter{\romannumeral #1\relax}}

\newcommand{\bv}{\ensuremath{\underline{b}}}

\newcommand{\ev}{\ensuremath{\mathbf{e}}}

\newcommand{\mv}{\ensuremath{\mathbf{m}}}

\newcommand{\qv}{\ensuremath{\mathbf{q}}}
\newcommand{\rv}{\ensuremath{\mathbf{r}}}
\newcommand{\sv}{\ensuremath{\mathbf{s}}}
\newcommand{\uv}{\ensuremath{\mathbf{u}}}
\newcommand{\vv}{\ensuremath{\mathbf{v}}}

\newcommand{\wv}{\ensuremath{\mathbf{w}}}
\newcommand{\xv}{\ensuremath{\mathbf{x}}}
\newcommand{\yv}{\ensuremath{\mathbf{y}}}
\newcommand{\zv}{\ensuremath{\mathbf{z}}}

\newcommand{\etav}{\ensuremath{\boldsymbol{\eta}}}
\newcommand{\zetav}{\ensuremath{\boldsymbol{\zeta}}}
\newcommand{\lambdav}{\ensuremath{\boldsymbol{\lambda}}}

\newcommand{\Am}{\ensuremath{\mathbf{A}}}
\newcommand{\Dm}{\ensuremath{\mathbf{D}}}

\newcommand{\Xv}{\vec{{X}}}

\def \Mc{M_{\mathrm{c}}}

\def \Nc{N_{\mathrm{c}}}

\def \Pc{P_{\mathrm{c}}}

\def \Bc{B_{\mathrm{c}}}

\def\Pr{\mathrm{Pr}}

\DeclareMathAlphabet{\mcl}{OMS}{cmsy}{m}{n}

\newcommand{\wh}{\widehat}

\newlength\tikzwidth
\newlength\tikzheight

\definecolor{mycolor1}{rgb}{0.63529,0.07843,0.18431}%
\definecolor{mycolor2}{rgb}{0.00000,0.44706,0.74118}%
\definecolor{mycolor3}{rgb}{0.00000,0.49804,0.00000}%
\definecolor{mycolor4}{rgb}{0.87059,0.49020,0.00000}%
\definecolor{mycolor5}{rgb}{0.00000,0.44700,0.74100}%
\definecolor{mycolor6}{rgb}{0.74902,0.00000,0.74902}%

\usepackage{amsmath}
\usepackage{amssymb}
\usepackage{amsthm}

\title{Coded Demixing for Unsourced Random Access}
\author{\IEEEauthorblockN{
Jamison R. Ebert, \emph{Student Member, IEEE},
Vamsi K. Amalladinne, \emph{Member, IEEE}, \\
Stefano Rini, \emph{Member, IEEE},
Jean-Francois Chamberland, \emph{Senior Member, IEEE}, \\
Krishna R. Narayanan, \emph{Fellow, IEEE} \\}
\thanks{
This material is based upon work supported, in part, by the National Science Foundation (NSF) under Grants CCF-2131106 and CNS-2148354, by Qualcomm Technologies, Inc., through their University Relations Program, and by the Ministry of Science and Technology (MOST) under Grant 110-2221-E-A49-052.

}}

\begin{document}
\maketitle

\begin{abstract}
Unsourced random access (URA) is a recently proposed multiple access paradigm tailored to the uplink channel of machine-type communication networks.
By exploiting a strong connection between URA and compressed sensing, the massive multiple access problem may be cast as a compressed sensing (CS) problem, albeit one in exceedingly large dimensions.
To efficiently handle the dimensionality of the problem, coded compressed sensing (CCS) has emerged as a pragmatic signal processing tool that, when applied to URA, offers good performance at low complexity.
While CCS is effective at recovering a signal that is sparse with respect to a single basis, it is unable to jointly recover signals that are sparse with respect to separate bases. 
In this article, the CCS framework is extended to the demixing setting, yielding a novel technique called coded demixing.
A generalized framework for coded demixing is presented and a low-complexity recovery algorithm based on approximate message passing (AMP) is developed.
Coded demixing is applied to heterogeneous multi-class URA networks and traditional single-class networks.
Its performance is analyzed and numerical simulations are presented to highlight the benefits of coded demixing. 

\end{abstract}


\section{Introduction}
\label{section:Introduction}

The number of unattended devices communicating over wireless networks is expected to increase drastically over the next decades.
These machine-type devices, unlike their human counterparts, tend to infrequently transmit very short payloads over the network medium.
The handling of such sporadic transmissions is highly inefficient under traditional uplink multiuser coordination processes and thus necessitates the design of novel data-link layer protocols. 

An increasingly popular paradigm for such machine-type communication is that of unsourced random access (URA), described by Polyanskiy~\cite{polyanskiy2017perspective}. 
In the URA formulation, only a small percentage of the total device population is envisioned to be actively communicating with the base station at a time and their messages are envisioned to be small, i.e., on the order of one hundred bits. 
Each active user simultaneously transmits its message over regularly scheduled time slots using a common codebook, thus eliminating the overhead and latency associated with multiuser coordination. 
The receiver is then tasked with producing an unordered list of transmitted messages.  
We note that, under this framework, the receiver does not know the origin of a message unless a unique identifier is embedded within the message itself.

The predominant performance criterion for URA schemes is the per-user probability of error (PUPE), or the probability that a transmitted message is not recovered by the receiver.
The number of active devices is typically assumed to be known at the access point, and the output list size is correspondingly constrained.
Since URA devices operate with short payloads, finite block length bounds provide better insight than asymptotic information-theoretic results in this context.
Polyanskiy~\cite{polyanskiy2017perspective}~offers finite block length achievability bounds based on random Gaussian coding and maximum likelihood (ML) decoding.
While these benchmarks show enviable performance, the ensuing schemes are not computationally tractable in scenarios of practical interest.  
Over the past few years, significant research effort has gone into developing pragmatic algorithms that approach Polyanskiy's benchmarks with manageable complexity \cite{ordentlich2017isit, vem2017user, pradhan2020sparse, marshakov2019polar, pradhan2019polar, ahmadi2021random, amalladinne2019coded, fengler2019sparcs, amalladinne2020unsourced, han2021sparse, calderbank2018chirrup}.   
We expound on several of these schemes below.

\subsection{Existing URA Schemes}

Ordentlich and Polyanskiy authored the first low-complexity scheme for URA in \cite{ordentlich2017isit}.  
Therein, they propose dividing the transmission period of $n$ channel uses into slots.
Each active user then randomly selects a slot to transmit its message over.  
A concatenated coding structure is employed wherein the inner linear code recovers the modulo-$2$ sum of transmitted codewords within a slot, and the outer BCH code recovers the constituent codewords.  
As long as fewer than $T$ users select a slot, decoding is feasible. 
In \cite{vem2017user}, Vem et al.\ retain the slotted structure proposed in \cite{ordentlich2017isit}, but transmit each message over multiple slots. 
Furthermore, a preamble section is carved off from every message; this initial portion is transmitted via compressed sensing (CS) and its content determines the repetition pattern for the corresponding payload.
The second part is then encoded using an LDPC code and decoded using a message-passing decoder within a slot.  
Successive interference cancellation (SIC) is leveraged to remove the contribution of decoded messages across slots.  
This latter scheme provides a sizeable performance improvement over \cite{ordentlich2017isit}.
The authors in \cite{pradhan2020sparse} introduce sparse interleave division multiple access (IDMA), which employs a signal sparse graph across all transmissions and results in a further performance improvement for systems with a large number of active devices.

In \cite{marshakov2019polar}, Marshakov et al.\ show that replacing LDPC codes with polar codes in $T$-fold irregular repetition slotted ALOHA (IRSA) can result in substantial performance benefits.
Their scheme outperforms sparse IDMA when the number of active users is low, i.e., less than $125$ or so.
Instead of using time-division for separating codewords, Pradhan et al.\ propose using random spreading in  \cite{pradhan2019polar}.  
There, the message is again broken into two parts where the first part is used to select a spreading sequence and the second part is encoded using a polar code.  
The polar-encoded codeword is spread using the selected spreading sequence.  
At the receiver, an energy detector is employed to identify active sequences and the codewords are recovered using a list decoder. 
In \cite{ahmadi2021random}, the polar coding scheme is improved significantly when the number of active users is large by allowing the set of active users to use different power levels.
In \cite{han2021sparse}, Han et al.\ present a sparse Kronecker product (SKP) based scheme wherein a user's message is encoded as the Kronecker product of a sparse vector and an FEC-encoded vector.
The sparse component enables the receiver to use CS techniques to perform multiuser detection and the FEC portion provides significant performance benefits.
An iterative algorithm is employed at the receiver.  
This latter scheme uniformly outperforms \cite{pradhan2019polar} and nearly achieves Polyanskiy's finite block length achievability bound when the number of active users is low.

While significant progress was being made with traditional channel coding schemes for URA, the connection between URA and compressed sensing was being explored further.
This connection is made by considering a simple bijection on each user's $B$-bit message. 
Specifically, each user's $B$-bit message $\uv \in \mathbb{F}_2^B$ may be transformed into a $1$-sparse vector $\xv \in \mathbb{F}_2^{2^B}$, where the single unity entry in $\xv$ is at index $\left[ \uv \right]_{2}$, which represents an integer under radix-$2$. 
After converting its payload to a sparse vector $\xv$, each user compresses its sparsified payload $\xv$ using a known sensing matrix $\Am \in \mathbb{R}^{n \times 2^B}$ into vector $\yv$ and then transmits $\yv$ over the Gaussian multiple access channel (GMAC). 
At the receiver, standard CS support recovery is performed to recover the index vector corresponding to the sent payload.
This scheme easily extends to the multiple access case: if $K$ users are active during a given time slot, the receiver simply performs $K$-sparse recovery and is able to recover all sent messages, provided that no two users select the same message to transmit.

A challenging aspect of the compressed sensing view for URA lies in the sheer dimensionality of the problem, often exceeding $2^{100}$ in practice.  
At such dimensions, it becomes impractical to utilize standard CS solvers.
To circumvent this complexity issue, Amalladinne et al.\ propose a divide-and-conquer approach wherein each information message is broken up into fragments which are linked together via a tree-based outer code \cite{amalladinne2019coded}. 
The coded message components are then converted into a sparse index representation, compressed using the common codebook $\Am$, transmitted over the AWGN channel, and individually recovered using standard CS techniques such as non-negative least-squares (NNLS) or LASSO.
After each section has been processed, a graph-based decoder connects recovered message fragments back together in a process called stitching \cite{amalladinne2019coded}. 
This scheme was enhanced by Fengler, Jung, and Caire in \cite{fengler2019sparcs} through the joint encoding of codeword sections at the transmitter and the application of approximate message passing (AMP) as a CS decoder at the receiver.  
The AMP framework was modified in \cite{amalladinne2020unsourced} by allowing for soft information to be dynamically shared between the inner CS decoder and the outer LDPC decoder, resulting in $1$~dB gain when $K = 100$. 
In \cite{ebert2020hybrid}, Ebert et al.\ show that the weak form of spatial coupling induced by the AMP denoiser in \cite{amalladinne2020unsourced} is sufficiently strong to harness the benefits of spatial coupling within the system, thus allowing for the independent encoding of codeword sections. 
This results in a roughly ${1}/{2}$ reduction in computational complexity while maintaining nearly the same level of performance.
The scheme presented in \cite{amalladinne2020unsourced} represents the state of the art in compressed sensing-based schemes for URA.
While this latter scheme does not outperform certain channel coding based approaches such as \cite{pradhan2019polar}, it offers much lower computational complexity.
An even lower-complexity CS-based scheme is described by Calderbank and Thompson in \cite{calderbank2018chirrup}; however, the complexity reduction comes at the expense of PUPE performance and it requires additional channel uses.
It should be noted that other URA schemes have been presented, e.g., \cite{kasper2020scheduling,decurninge2020tensorbased,shyianov2020massive,nassaji2022unsourced}, yet they are only peripherally connected to our contributions.
Indeed, the ideas presented in this work extend some of the notions introduced by Amalladinne et al.~\cite{amalladinne2020unsourced} and strengthen the connection between URA and the field of sparse recovery.

\subsection{Main Contributions}

From a signal processing perspective, one reason for the recent interest in CCS is that it is a computationally efficient tool for recovering signals of exceedingly high dimensions (on the order of $2^{100}$) that are sparse in a given domain.
Given the connection between URA and compressed sensing, CCS is a natural solution to the massive random access problem. 

Independent of CCS, demixing (or convex demixing) has emerged as a tool for recovering constituent signals $\sv_1$ and $\sv_2$ from their sum $\sv_1 + \sv_2$, given that $\sv_1$ and $\sv_2$ are sufficiently sparse in incoherent bases \cite{mccoy2014convexity}. 
In this paper, we draw connections between CCS and demixing to allow for the efficient recovery of very high dimensional signals that are sparse in different bases, where the separate bases exhibit low cross-coherence. % provided that technical conditions on sparsity and incoherence are satisfied.
Due to the extreme dimensions of the problem, the divide and conquer approach of CCS is extended to the demixing problem, thus yielding a novel framework that we call \textit{coded demixing}. 

When applied to URA, coded demixing provides the best known performance to  multiple heterogeneous classes of users that simultaneously utilize limited channel resources.
Furthermore, coded demixing provides significant gains over CCS in the single-class network by reducing the number of users present on a single outer factor graph.
We summarize our main contributions as follows.

\begin{enumerate}
    \item In Section~\ref{section:CodedDemixing}, a framework for coded demixing is carefully developed that allows for the compression and simultaneous transmission of multiple sparse signals from incoherent bases over fixed channel resources.
    At the receiver, a low-complexity AMP-based recovery algorithm performs demixing at scale.
    \item In Section~\ref{section:MulticlassviaCodedDemixing}, the developed coded demixing framework is applied to a network containing several classes of devices, where a class of devices is a collection of devices with a fixed message length, power budget, and set of coding requirements. 
    Numerical simulations are presented to show that, in the multi-class context, coded demixing outperforms treating interference as noise (TIN) and successive interference cancellation (SIC) schemes with comparable computational complexity.
    \item In Section~\ref{section:StochasticBinningandCodedDemixing}, the developed coding demixing framework is applied to a network consisting of a single class of users.
    There, the set of active users is randomly partitioned into distinct, yet homogeneous, groups during message transmission. 
    Indeed, this reduces the expected number of users present on a single factor graph and therefore improves the performance of the belief propagation (BP) decoder. 
    It is shown that coded demixing with stochastic binning provides nearly state-of-the-art performance when the number of users is large in the single-antenna URA network at reduced complexity.
\end{enumerate}

\subsection{Notation}
Unless specifically noted, all variables are real-valued.  
Matrices are denoted by bold capital letters such as $\Am$, and column vectors are denoted by bold lower case letters such as $\xv$. 
The transpose of $\xv$ is denoted by $\xv^\intercal$ and the $\ell_p$-norm of $\xv$ is denoted $\| \xv \|_{p}$. 
Furthermore, the concatenation of two vectors is written as $\uv\vv$, where $\uv\vv \triangleq [\uv^\intercal \vv^\intercal]^\intercal$. 
An $n \times n$ diagonal matrix with entries $d_1, d_2, \hdots d_n$ along its main diagonal is denoted as $\operatorname{diag}\left(d_1, d_2, \hdots d_n\right)$.
In general, sets are denoted by calligraphic letters such as $\mathcal{G}$; two exceptions to this rule are $\mathbb{R}$, which denotes the set of real numbers; and $[N]$, which denotes the set of integers $\{ 1, 2, \ldots, N \}$. 
The cardinality of a set is denoted by $|\mathcal{G}|$.
Finally, we denote the probability of event $\mathcal{X}$ by $\mathbb{P}\left(\mathcal{X}\right)$.

\section{System Model and Coded Demixing}
\label{section:CodedDemixing}

In this section, we present the system under consideration and introduce coded demixing both as a conceptual framework and as a solution to various URA scenarios.
From an abstract point of view, coded demixing extends the CCS family of algorithms~\cite{amalladinne2019coded,amalladinne2020unsourced,fengler2019sparcs} to situations where the received signal is the superposition of multiple components that are known to be sparse in distinct domains.
The presence of multiple sparse domains introduces new challenges including the eventual need to estimate the number of messages in each domain and the construction of an efficient recovery scheme tailored to this more general framework.
In doing so, we leverage several notions from CCS and, whenever appropriate, we draw distinctions about the unique character of coded demixing compared to prior art.
Initially, we consider a case where side information is available in terms of number of active devices per sparse domain; this assumption will be relaxed in Sec.~\ref{section:StochasticBinningandCodedDemixing}.

\subsection{System Model}
\label{section:SystemModel}

Consider a URA system in which $K$ active devices out of $K_{\mathrm{tot}}$ total devices wish to communicate with a base station over a single-input single-output (SISO) uncoordinated GMAC.
Each of the $K$ active devices belongs to one of $G$ groups; the manner in which these groups are created will be described later in this article. 
Devices within a group are homogeneous, but differences in power, data, and/or coding requirements may exist across groups.
For convenience, we assign a unique label $j \in [K]$ to every currently active device.
Although these labels appear in some equations, they are immaterial to exchanges between devices and the access point.
Furthermore, we emphasize that these generic labels are employed solely for the purpose of exposition; they do not reveal anything about the true identities of the active users.
The notation $\mathcal{G}_g$ refers to the set of active users that belong to group $g \in [G]$ and $K_g = |\mathcal{G}_g|$ denotes the number of active users in that group.
User~$j$ encodes its message $\wv_j$ into the signal $\xv_j \in \mathbb{R}^{n}$ and then transmits $\xv_j$ over $n$ channel uses.
The signal received at the access point is given by
\begin{equation} \label{equation:URA_coded_demixing}
    \yv = \sum_{g \in [G]}\sum_{j \in \mathcal{G}_g}d_g\xv_j + \zv,
\end{equation}
where $d_g$ represents the amplitude scaling within group $g$ and $\zv \in \mathbb{R}^n$ is additive white Gaussian noise with i.i.d.\ standard normal components. 
The signal sent by user $j \in \mathcal{G}_g$, denoted $\xv_j$, is produced via a codebook $\mathcal{C}_g$ common to all users in $\mathcal{G}_g$.
We emphasize that the transmitted signal is only a function of the user's data and group, and \textit{not} its identity. 
It is also noted that the signal $\xv_j \in \mathcal{C}_g \subset \mathbb{R}^n$ satisfies the URA power constraint $\mathbb{E} \left[ \| \xv_j \|_2^2 \right] \leq nP_g$. 

The receiver is then tasked with recovering an unordered list of transmitted messages $\hat{\mathcal{W}}(\yv)$, where $|\hat{\mathcal{W}}(\yv)| \leq K$. 
The performance of the scheme is evaluated via the per user probability of error $P_e$,
\begin{equation}
    \label{eq:pupe}
    P_e = \frac{1}{K}\sum_{j \in [K]} \mathbb{P} 
    \lb \wv_j \notin \hat{\mathcal{W}}(\yv) \rb.
    %\left\{ \wv_j \notin \hat{\mathcal{W}}(\yv) \right\} .
\end{equation}
As is customary in existing URA literature, the total number of active users $K$ is given as side information at the receiver.
In contrast, the number of active users contained within each group is unknown and therefore must be estimated.

Aside from the presence of multiple groups, this system model is equivalent to the model adopted in other URA articles when the access point features a single receive antenna.
This enables the fair comparison of our proposed scheme with prior art, using a \emph{de facto} common task framework.

\subsection{From CCS to Coded Demixing}
\label{section:CCS approach}

The theory of coded demixing introduced in this article is closely connected to CCS, and especially the variant of CCS found in \cite{amalladinne2020unsourced}.
Still, the presence of multiple sparse domains prevents the direct application of CCS to the URA formulation in \eqref{equation:URA_coded_demixing}.
Below, we discuss the innovations needed to enable coded demixing, focusing primarily on the message encoding and the decoding processes.
For the time being, we assume that the values of $\{ K_g \}$ are known; we will eventually circle back and relax this requirement.
We delay applications and performance characterization to subsequent sections (see Sec.~\ref{section:MulticlassviaCodedDemixing} and Sec.~\ref{section:StochasticBinningandCodedDemixing}).

\subsubsection{Encoding Process}

Without loss of generality, we present the encoding process for user~$j$ in group~$g$.  
We note that the structure of the encoding process is the same across groups, but specific parameters may vary from group to group. 
The message encoding process utilizes the concatenated coding structure established in \cite{amalladinne2019coded}.
This features a SPARC-like CS inner code similar to the approach proposed in \cite{fengler2019sparcs}, and an LDPC outer code akin to those in found~\cite{amalladinne2020unsourced}.

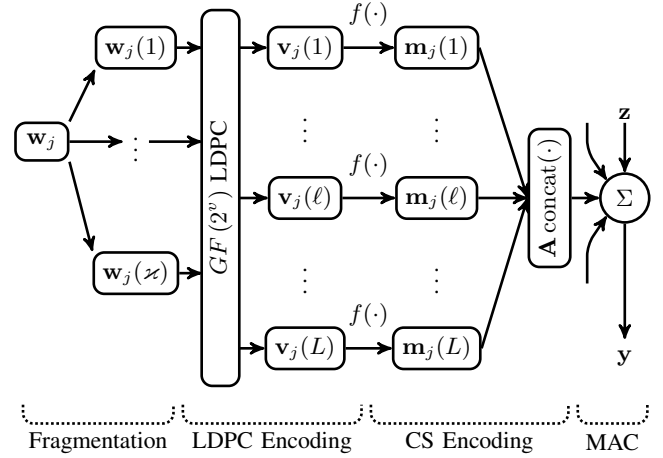
\begin{figure}
\centering
\begin{tikzpicture}
    [
    font=\small, >=stealth', line width=1pt, draw=black,
    block/.style={rectangle, draw, minimum height=5mm, minimum width=7mm, rounded corners},
    section/.style={circle, minimum size=7mm, draw=black},
    check/.style={rectangle, minimum height=6mm, minimum width=6mm, draw=black, fill=gray!20, rounded corners},
    trivialcheck/.style={rectangle, minimum height=4mm, minimum width=4mm, draw=black, fill=gray!20},
    section/.style={circle, minimum size=7mm, draw=black},
    ]

% Stage 1: user's starting message block
\node[block](wj) at (0.25, 3.25) {$\mathbf{w}_j$};

% Stage 2: user's message fragments
\node[block](wj1) at (1.5, 4.5) {$\mathbf{w}_j(1)$};
\node[draw=none](wjdots1) at (1.5, 3.25) {$\vdots$};
\node[block](wjkappa) at (1.5, 1.5) {$\mathbf{w}_j(\varkappa)$};

% Stage 1-2 connections
\draw[->, rounded corners] (wj.north east) -- (wj1.south west);
\draw[->] (wj.east) -- (wjdots1.west);
\draw[->] (wj.south east) -- (wjkappa.north west);

% Stage 3: Non-binary LDPC Encoder
\draw[draw=black, fill=white, rounded corners] (2.375, 0) rectangle (2.875, 5);
\node[draw=none, rotate=90] (ldpclabel) at (2.625, 2.5) {$GF\left(2^{v}\right)$ LDPC};

% Stage 2-3 connections
\draw[->] (wj1.east) -- (2.375, 4.5);
\draw[->] (wjdots1.east) -- (2.375, 3.25);
\draw[->] (wjkappa.east) -- (2.375, 1.5);

% Stage 4: Codeword fragments
\node[block](vj1) at (3.75, 4.5) {$\mathbf{v}_j(1)$};
\node[draw=none](vjdots1) at (3.75, 3.5) {$\vdots$};
\node[block](vjell) at (3.75, 2.5) {$\mathbf{v}_j(\ell)$};
\node[draw=none](vjdots2) at (3.75, 1.5) {$\vdots$};
\node[block](vjL) at (3.75, 0.5) {$\mathbf{v}_j(L)$};

% Stage 3-4 connections
\draw[->] (2.875, 4.5) -- (vj1.west);
\draw[->] (2.875, 2.5) -- (vjell.west);
\draw[->] (2.875, 0.5) -- (vjL.west);

% Stage 5: sparse message fragments
\node[block](mj1) at (5.5, 4.5) {$\mathbf{m}_j(1)$};
\node[draw=none](mjdots1) at (5.5, 3.5) {$\vdots$};
\node[block](mjell) at (5.5, 2.5) {$\mathbf{m}_j(\ell)$};
\node[draw=none](mjdots2) at (5.5, 1.5) {$\vdots$};
\node[block](mjL) at (5.5, 0.5) {$\mathbf{m}_j(L)$};

% Stage 4-5 connections
\draw[->] (vj1.east) -- node[above,yshift=1.5mm] {$f(\cdot)$} (mj1.west);
% \node[draw=none] (flabel1) at (4.45, 4.8) {$f(\cdot)$};
\draw[->] (vjell.east) -- node[above,yshift=1.5mm] {$f(\cdot)$} (mjell.west);
% \node[draw=none] (flabelell) at (4.45, 2.8) {$f(\cdot)$};
\draw[->] (vjL.east) -- node[above,yshift=1.5mm] {$f(\cdot)$} (mjL.west);
% \node[draw=none] (flabelL) at (4.45, 0.8) {$f(\cdot)$};

% Stage 6: CS compression
\node[block, rotate=90] (encode) at (7.0, 2.5) {$\mathbf{A}\operatorname{concat}(\cdot)$};

% Stage 5-6 connections
\draw[->] (mj1.east) -- (encode.north);
\draw[->] (mjell.east) -- (encode.north);
\draw[->] (mjL.east) -- (encode.north);

% Stage 7 MAC channel
\node[draw, circle] (MAC) at (8, 2.5) {$\Sigma$};
% \node[draw, circle] (MAC) at (8.5, 3.375) {$\Sigma$}; Other

% % Stage 6-7 connections
% \draw[->] (encode.south) -- (MAC.west);
% \node[draw=none] (xj) at (7.5, 2.75) {$\mathbf{x}_j$};
% \draw[->] (8, 3.5) -- (MAC.north);
% % \node[draw=none] (awgn) at (8, 3.75) {$\mathbf{z}$};
% \node[draw=none] (awgn) at (8, 3.625) {$\mathbf{z}$};
% \draw[->] (7.75, 1.5) -- (MAC);
% \node[draw=none] (mjp) at (7.50, 1.25) {$\mathbf{x}_{\Acute{j}}$};
% \draw[->] (8.25, 1.5) -- (MAC);
% \node[draw=none] (mjpp) at (8.50, 1.25) {$\mathbf{x}_{\Tilde{j}}$};
% \draw[->] (MAC.east) -- (8.75, 2.5);
% \node[draw=none] (output) at (8.625, 2.75) {$\mathbf{y}$};

% Stage 5-6 connections
\draw[->] (8, 3.5) -- (MAC.north);
\node[draw=none] (awgn) at (8, 3.625) {$\mathbf{z}$};
\draw[->, rounded corners] (7.5, 1.3625) -- (7.5, 1.875) -- (MAC.south west);
\draw[->, rounded corners] (7.5, 3.5) -- (7.5, 3.125) --  (MAC.north west);

% \draw[->] (MAC.south) -- (8.5, 1.5);
% \node[draw=none] (ylabel) at (8.5, 1.25) {$\mathbf{y}$};

\draw[->] (encode.south) -- (MAC.west);
% \node[draw=none] at (7.5, 2.625) {$\xv_j$};
% \draw[->] (8.5, 3.25) -- (MAC.north);
% \draw[->, rounded corners] (7.5, 1.1125) -- (7.5, 1.625) -- (MAC.south west);
% \draw[->, rounded corners] (7.5, 3.25) -- (7.5, 2.875) --  (MAC.north west);
\draw[->] (MAC.south) -- (8, 0.625);
\node[draw=none] (ylabel) at (8, 0.375) {$\mathbf{y}$};

% Fancy labels?
\draw[densely dotted, rounded corners] (0, -0.25) -- (0, -0.5) -- (2, -0.5) -- (2, -0.25);
\node[draw=none] (fragmentationlabel) at (1, -0.75) {Fragmentation};
\draw[densely dotted, rounded corners] (2.125, -0.25) -- (2.125, -0.5) -- (4.5, -0.5) -- (4.5, -0.25);
\node[draw=none] (ldpcencodinglabel) at (3.3125, -0.75) {LDPC Encoding};
\draw[densely dotted, rounded corners] (4.625, -0.25) -- (4.625, -0.5) -- (7.25, -0.5) -- (7.25, -0.25);
\node[draw=none] (csencodinglabel) at (5.9375, -0.75) {CS Encoding};
\draw[densely dotted, rounded corners] (7.375, -0.25) -- (7.375, -0.5) -- (8.30, -0.5) -- (8.30, -0.25);
\node[draw=none] (macchannellabel) at (7.8375, -0.75) {MAC};
    
\end{tikzpicture}
\caption{
    This figure illustrates the encoding process for user $j$ in group $g$. 
    For simplicity of notation, the group subscript $g$ has been dropped in this figure. 
    User $j$'s message $\wv_j$ is first broken into $\varkappa_g$ fragments; then, these fragments are encoded using an LDPC code over $GF\left(2^{v_g}\right)$. 
    Each of the LDPC-encoded fragments are subsequently converted into $1$-sparse vectors through the bijection $f_g\left(\cdot\right)$ and the resulting sparse vectors are compressed through multiplication with $\Am_g$.
    This results in signal $\xv_j$, which is combined with signals from other users (possibly from other groups) and AWGN to form the received signal $\yv$.
}
\label{figure:EncodingProcess}
\end{figure}

The first step in the encoding process is to add redundancy to user~$j$'s $w_g$-bit message $\wv_j$ via a rate $R_g$ LDPC code.
Explicitly, the message $\wv_j \in \{0, 1\}^{w_g}$ is broken into $\varkappa_g$ segments each of length $v_g$ such that
\begin{equation*}
    \wv_j = \wv_j(1)\wv_j(2)\cdots\wv_j(\varkappa_g).
\end{equation*}
Each section $\wv_j(i)$ may be viewed as an element of GF($2^{v_g}$) and thus the entire message $\wv_j$ may be encoded using a non-binary LDPC code, yielding a codeword $\vv_j$. 
The codeword $\vv_j$ can likewise be broken into $L_g$ segments, each of length $v_g$:
\begin{equation} \label{equation:codeword}
    \vv_j = \vv_j(1)\vv_j(2)\cdots\vv_j(L_g) .
\end{equation}
At this point, each section of the codeword is converted into a $1$-sparse message vector via the bijection $\mv_j(i) = f_g(\vv_j(i))$, where $f_g(\vv(i)): \mathbb{R}^{v_g} \rightarrow \mathbb{R}^{2^{v_g}}$. 
The single unitary entry of $\mv_j(i)$ is at location $[\vv_j(i)]_2$, which is interpreted as an integer expressed under radix-$2$.
Once every section is encoded, the resulting $L_g$-sparse vector $\mv_j = \mv_j(1)\cdots\mv_j(L_g)$ becomes similar to a SPARC codeword; although it features fewer sections, each is of much greater length \cite{fengler2019sparcs}.
Attached to group~$g$ is a Gaussian sensing matrix $\Am_g \in \mathbb{R}^{n \times L_g 2^{v_g}}$ with independent entries distributed as $\mathcal{N}(0, {1}/{n})$, so that the expected $\ell_2$-norm of each column is one.
Message $\mv_j$ is converted into a signal $\xv_j = \Am_g\mv_j$ and transmitted over the channel.  
Fig.~\ref{figure:EncodingProcess} graphically summarizes this part of the encoding process.
From the perspective of a device, these steps are equivalent to the encoding process within CCS.

The distinction between CCS and coded demixing becomes more apparent when the distributed encoding of all active devices is considered together.
We let vector $\sv_g$, $g \in [G]$, be the sum of all the signals sent by active devices within group~$g$.
That is, 
\begin{equation*}
    \sv_g = \sum_{j \in \mathcal{G}_g} \mv_j .
\end{equation*}
We allow devices across groups to use different power levels, which leads to the scaling of signal amplitudes, with $\Phi_g = d_g \Am_g$.
Under these aggregates, the signal received at the base station can be expressed as
\begin{equation}
    \label{eq:demixingform}
    \yv = \Phi_1\sv_1 + \Phi_2\sv_2 + \cdots + \Phi_G\sv_G + \zv,
\end{equation}
where $\sv_g$ is $L_g K_g$-sparse.
The random nature of each $\Am_g$ ensures that the set of sensing matrices $\{\Am_g : g\in[G]\}$ exhibits low cross-coherence with high probability; this is key for the successful application of demixing techniques \cite{mccoy2014sharp}. 
With the conditions of sparsity and low coherence met, \eqref{eq:demixingform} assumes a canonical form for demixing problems.

\begin{remark}
\label{remark:cvxoptcomplexity}
The sparse domains in coded demixing are defined by the column spans of $\{\Am_g : g\in[G]\}$.
Each group typically uses a separate LDPC outer code, and the number of sections and their lengths may vary from group to group.
The introduction of multiple groups creates the need to track $\{ K_g : g \in [G] \}$; these quantities may not be available as side information.
This, in turn, may require the modification of the transmitted signals, with possibly the addition of group specific pilots, to facilitate this task.
\end{remark}

\begin{remark}
The presence of multiple sparse domains invites the rethinking of the decoder because the naive application of parallel CCS decoders (one for each sparse domain) to the received vector $\yv$ will not work.
Intuition can be drawn from the many algorithms designed for convex demixing~\cite{mccoy2014convexity, mccoy2014sharp, amelunxen2014living, Chandrasekaran_2012, hedge2012signal, bhaskar2013atomic}, yet a suitable demixing solver must have manageable complexity for the problem dimensions relevant to modern communication scenarios and should take advantage of the signal structure described above.
These requirements preclude the use of classical regularized convex optimization approaches; consequently, we resort to a first-order iterative solution. 
\end{remark}

\subsubsection{Decoding Process}

\begin{figure*}[t]
\centering
\begin{tikzpicture}
  [
  font=\small, >=stealth', line width=1pt,
  check/.style={rectangle, minimum height=2.5mm, minimum width=2.5mm, draw=black},
  varnode/.style={circle, minimum size=2mm, draw=black},
  mmse/.style={rectangle, minimum height=7.5mm, minimum width=25mm, rounded corners, draw=black},
  quantity/.style={rectangle, minimum height=8mm, minimum width=8mm, rounded corners, draw=black},
  multiply/.style={trapezium, trapezium angle=75, draw=black, minimum width=10mm, minimum height=8mm, rounded corners}
  ]

% Received Signal
%
\node[quantity, label=Received Signal] (signal) at (-7,3.5) {$\mathbf{y}$};
\draw[*->, rounded corners] (-9.25, 4.5) -- (-9.25, 3.5) -- node[below] {Input} (signal);

% Residual
%
\node[quantity] (residual) at (-5.5,2) {$\mathbf{z}^{(t)}$};
\node[circle, minimum width=8mm, draw=black] (sum) at (-7,2) {$\sum$}
  edge[<-] (signal)
  edge[->] (residual);
\node[quantity] (stack) at (-5,-2) {Stack};
\draw[*->, rounded corners] (stack) -- (-5,-4) -- node[above] {Output} (-3,-4);
\node[multiply,label=below:Amplitude] (Dmatrix) at (-7,-2) {$\Dm$}
  edge[<-] (stack);
\node[multiply] (Amatrix) at (-7,0.5) {$\Am$}
  edge[->] node[right] {$-$} (sum)
  edge[<-] (Dmatrix);

% % % % % % % % % %

\foreach \g/\xg/\yg in {3/3.5/2,2/2.25/1,1/1/0} {
    % Effective Observation
    %
    \draw[draw=black, densely dotted, rounded corners, fill=white] (-4.25+\xg,-0.25+\yg) rectangle (0.375+\xg,2.75+\yg);
    % \node[draw=none] (effectiveobservation\g) at (-1.9375+\xg,3+\yg) {Effective Observation};
    \node[multiply, shape border rotate=270] (dual\g) at (-3.5+\xg,2+\yg) {$\Am_\g^\intercal$};
    \draw[->, rounded corners] (residual.east) -- (-5.5+\xg,2+\yg) -- (dual\g);
    \node[circle, minimum width=8mm, draw=black] (sum\g) at (-2+\xg,2+\yg) {$\sum$}
      edge[<-] (dual\g);
    \node[quantity] (rv\g) at (-0.5+\xg,2+\yg) {$\rv_\g^{(t)}$}
      edge[<-] (sum\g)
      edge[->] (1+\xg,2+\yg);
    \node[multiply] (amplitude\g) at (-2+\xg,0.5+\yg) {$\Dm_\g$}
      edge[->] (sum\g);
    
    % Denoiser
    %
    \draw[draw=black, densely dotted, rounded corners, fill=white] (1+\xg,-0.75+\yg) rectangle (4+\xg,2.75+\yg);
    % \node[draw=none] (denoiser\g) at (2.5+\xg,3+\yg) {Denoiser};
    \node[mmse] (mmse\g) at (2.5+\xg,-0.125+\yg){Dynamic PME};
    \draw[draw=black, rounded corners] (1.25+\xg,1+\yg) rectangle (3.75+\xg,2.5+\yg);
    \foreach \s in {1,2,3,4,5} {
      \node[varnode] (var\g-\s) at (1+0.5*\s+\xg,1.25+\yg) {}
        edge (mmse\g);
    }
    \foreach \c in {1,2,3} {
      \node[check] (check\g-\c) at (1+0.75*\c+\xg,2.25+\yg) {};
    }
    \draw (var\g-5) -- (check\g-3.south);
    \draw (var\g-4) -- (check\g-3.south);
    \draw (var\g-3) -- (check\g-3.south);
    \draw (var\g-2) -- (check\g-2.south);
    \draw (var\g-5) -- (check\g-2.south);
    \draw (var\g-4) -- (check\g-1.south);
    \draw (var\g-2) -- (check\g-1.south);
    \draw (var\g-1) -- (check\g-1.south);
    
    % State Update
    %
    \node[quantity] (state\g) at (-2+\xg,-3+\yg) {$\sv_\g^{(t)}$}
      edge[->] (amplitude\g);
    \draw[->,rounded corners] (state\g) -- (-4+\xg,-3+\yg) -- (stack.east);
    \node[quantity] (state\g-delay) at (2.5+\xg,-3+\yg) {Delay}
      edge[<-] (2.5+\xg,-0.75+\yg)
      edge[->] (state\g);
}

% % % % % % % % % %

% Onsager Term
%
\draw[draw=black, densely dotted, rounded corners] (-10,-2.75) rectangle (-8.0, 0.25);
\node[draw=none,rotate=90] (onsagerlabel) at (-10.25, -1.25) {Onsager Term};
\node[quantity] (onsagerdelay) at (-5.5,-0.5) {Delay}
  edge(residual);
\node[quantity] (div) at (-9,-2) {$\frac{1}{n} \mathrm{div} (\cdot)$}
  edge[<-,dashed] (Dmatrix);
\node[circle, minimum width=8mm, draw=black] (times) at (-9,-0.5) {$\times$}
  edge[<-,dashed] (div)
  edge[<-,dashed] (onsagerdelay);
\draw[->,dashed,rounded corners] (times) -- (-9,2) -- (sum);

\end{tikzpicture}
\caption{
    This notional diagram depicts the operation of the coded demixing recovery algorithm.
    The input comes in the form of observation $\yv$ at the top left.
    The contributions of the state estimate (minus the Onsager term) are subtracted from $\yv$, leading to the residual $\zv$.
    This residual is then processed through demixing and turned into an effective observation for each group.
    Denoising is performed as a means to update the state estimates.
    The computation of the Onsager contribution, which is intrinsic to AMP, is highlighted using dashed lines.
    This iterative process continues until convergence is reached.
    At that point, the state estimates $\{\sv_g : g \in [G]\}$ are taken as the output of the algorithm.
}
\label{figure:DecodingProcess}
\end{figure*}
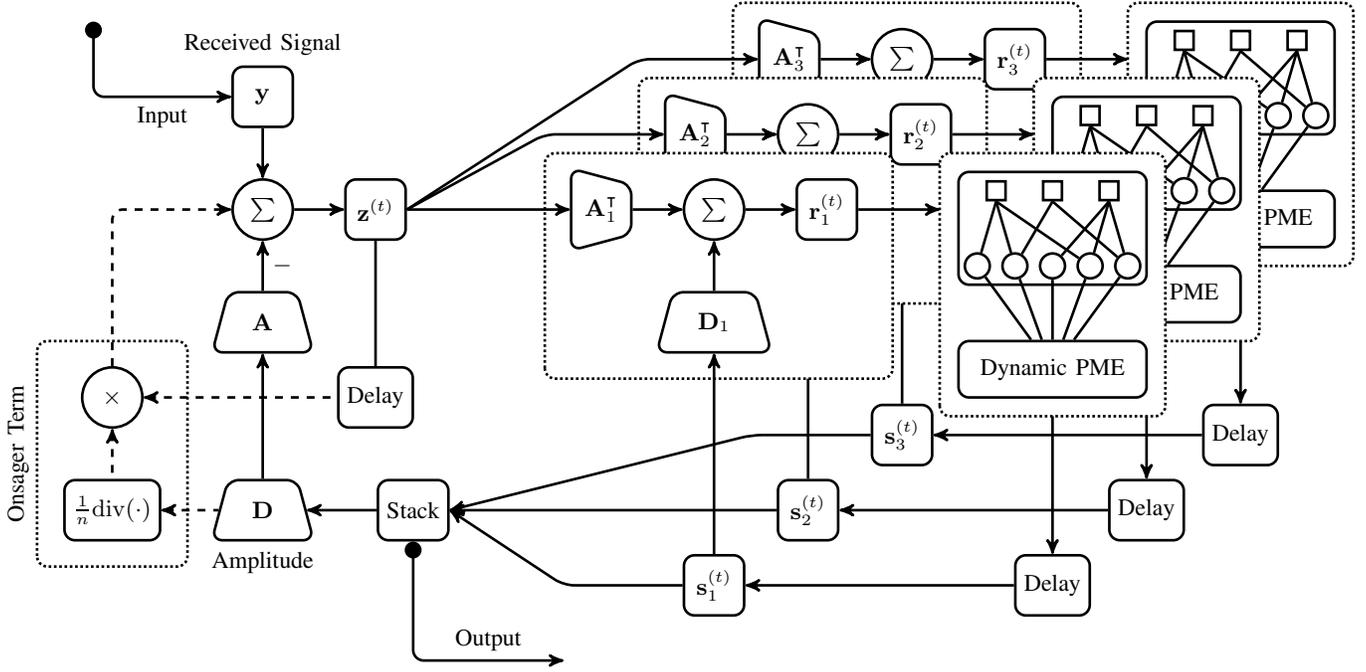

The decoder is tasked with recovering the set $\{\sv_g : g\in[G]\}$ and subsequently, disambiguating the messages contained within each $\sv_g$.  
This two-step process is inspired by CCS decoding as presented in \cite{amalladinne2020unsourced}, albeit with key modifications to support multiple groups of users.
A diagram of the iterative decoding procedure, with its main components, appears in Fig.~\ref{figure:DecodingProcess}.

Equation \eqref{eq:demixingform} admits an alternate representation that gives insight into how to develop an iterative procedure for support recovery.
If the matrices $\{\Phi_g : g \in [G]\}$ are concatenated to form $\Phi = [\Phi_1, \Phi_2, \ldots, \Phi_G] \in \mathbb{R}^{n \times \gamma}$, where $\gamma = \sum_g L_g 2^{v_g}$, and the vectors $\{\sv_g : g\in [G]\}$ are similarly stacked into $\sv = \sv_1 \sv_2 \ldots \sv_G \in \mathbb{R}^{\gamma}$, then \eqref{eq:demixingform} can be rewritten as
\begin{equation}
    \label{eq:demixingampform}
    \yv = \Phi\sv + \zv = \Am \Dm \begin{bmatrix}\sv_1 \\ \vdots \\ \sv_G\\ \end{bmatrix} + \zv,
\end{equation}
where $\Dm = \operatorname{diag}(d_1, d_2, \ldots, d_G)$ and $\Am = [\Am_1, \Am_2, \ldots, \Am_G]$.
We note that \eqref{eq:demixingampform} is in standard CS form, and possibly amenable to decoding via approximate message passing (AMP).
Of course, the rewriting of this equation does not simplify the intricate sparse structure of vectors $\{ \sv_g : g \in [G] \}$.
However, it offers a conceptual bridge that has been leveraged in the past to relate demixing to CS.
At this point, we turn to the AMP composite iterative structure and to the design of suitable denoiser for coded demixing.
The specific AMP algorithm we have in mind takes the form
\begin{gather}
    \label{eq:ampresidual}
    \zv^{(t)} = \yv - \Am \Dm \begin{bmatrix}\sv_1^{(t)} \\ \vdots \\ \sv_G^{(t)}\\ \end{bmatrix}
    + \frac{\zv^{(t - 1)}}{n}\mathrm{div}\Dm\etav^{(t-1)}\left(\rv^{(t-1)}\right) \\
    \label{eq:ampstateupdate}
    \begin{bmatrix} \sv_1^{(t+1)} \\ \vdots \\ \sv_G^{(t+1)}\\ \end{bmatrix}
    = \etav^{(t)}\left(\rv^{(t)}\right) = \begin{bmatrix} \etav_1^{(t)}\left(\rv_1^{(t)}\right) \\
    \vdots \\ \etav_G^{(t)}\left(\rv_G^{(t)}\right) \\
    \end{bmatrix},
\end{gather}
where the superscript $t$ denotes the iteration number, $\sv^{(0)} = \mathbf{0}$, and $\zv^{(0)} = \yv$.
Although more involved, this is the natural multi-group extension of the AMP algorithm applied to CCS in~\cite{amalladinne2020unsourced}.
In the present context, we define the effective observation by
\begin{equation}
    \label{eq:ampeffectiveobservation}
    \rv^{(t)} = \Am^\intercal \zv^{(t)}
    + \Dm \begin{bmatrix} \sv_1^{(t)} \\ \vdots \\ \sv_G^{(t)} \\ \end{bmatrix}.
\end{equation}
Equation~\eqref{eq:ampresidual} can be interpreted as the residual enhanced with an Onsager correction term~\cite{bayati2011dynamics, donoho2013information}.
The AMP state is updated in \eqref{eq:ampstateupdate} using denoiser $\eta_g^{(t)}(\cdot)$, which seeks to take advantage of the block structure of the problem.
Essentially, the denoiser should promote structured sparsity while also accounting for the presence of the outer code and, as such, it must be designed carefully.
The two iterative steps are linked through the computation of the effective observation in \eqref{eq:ampeffectiveobservation}.

A crucial fact about AMP is that under suitable conditions, such as $\Av$ having i.i.d.\ Gaussian entries and the denoiser being Lipschitz-continuous \cite{bayati2011dynamics,berthier2020state}, $\rv^{(t)}$ is asymptotically distributed as
\begin{equation}
    \label{eq:asymptoticeffectiveobservation}
    \Dm \begin{bmatrix} \sv_1^{(t)} \\ \vdots \\ \sv_G^{(t)} \\ \end{bmatrix}
     + \tau_t \zetav_t,
\end{equation}
where $\zetav_t \sim \mathcal{N}(0, \mathbf{I})$ and $\tau_t$ is a deterministic scalar quantity.
It is also noted that, though $\tau_t$ may be computed through joint state evolution, it is often approximated by $\tau_t^2 \approx \| \zv^{(t)} \|_2^2 / n$. 
Naturally, it may be tempting to apply a minimum mean square error (MMSE) estimator to each section $\ell$
\begin{equation*}
    \mathbb{E}[\sv_g(\ell) | d_g\sv_g + \tau_t\zetav_t = \rv_g]    
\end{equation*}
as the denoiser.
However, because each $\sv_g(\ell)$ is $K_g$-sparse, the MMSE estimator would have to consider $\binom{2^{v_g}}{K_g}$ different possibilities for every section.
Clearly, this becomes combinatorically intractable for section sizes of practical interest.
We therefore adopt the strategy proposed by Fengler et al.~\cite{fengler2019sparcs},  and we approximate the MMSE denoiser using the posterior mean estimator (PME),
\begin{equation} \label{eq:OriginalPME}
    \begin{split}
        \hat{s}_{g} \left( q, r, \tau_t \right) &=     \\
        &\frac{q \exp \left( - \frac{ \left( r - d_{g} \right)^2}{2 \tau_t^2} \right)}
        { q \exp \left( - \frac{ \left( r -  d_{g} \right)^2}{2 \tau_t^2} \right)
        + (1-q) \exp \left( -\frac{r^2}{2 \tau_t^2} \right)}, \\
    \end{split}
\end{equation}
where $r$ is an entry in the effective observation $\rv$, $q$ denotes the prior probability of an entry of $\sv_{g}$ being equal to one, and $\tau_t$ is the scaling factor discussed above.
In \cite{fengler2019sparcs}, the authors focus on the single-group scenario and they employ an uninformative prior, which would be equivalent to $q_g = 1 - (1 - 2^{-v_g})^{K_g}$ for group~$g$.

\begin{figure}[t!]
    \centering
    \begin{tikzpicture}
  [
  font=\small, >=stealth', line width=1pt, draw=black,
  check/.style={rectangle, minimum height=6mm, minimum width=6mm, draw=black, fill=gray!20},
  trivialcheck/.style={rectangle, minimum height=4mm, minimum width=4mm, draw=black, fill=gray!20},
  section/.style={circle, minimum size=7mm, draw=black},
  trivialcheck/.style={rectangle, minimum height=4mm, minimum width=4mm},
  pmeblock/.style={rectangle, minimum height=6mm, minimum width=2mm},
  quantity/.style={rectangle, minimum height=9mm, minimum width=20mm, rounded corners, draw=black, fill=white, line width=1pt},
  psbox/.style={rectangle, minimum height=6mm, minimum width=75mm, rounded corners, draw=black, fill=white, line width=1pt}
  ]
  
% Draw appropriate boxes
\draw[draw=black, fill=white, densely dotted, rounded corners] (-4.0,-1.125) rectangle (4.125, 2.25);
\node[draw=none, rotate=-90](fglabel) at (4.375, 0.625) {LDPC Factor Graph};
\draw[draw=black, fill=white, densely dotted, rounded corners] (-4.0, -2.125) rectangle (4.125, -3.875);
\node[draw=none, rotate=-90](overallpmelabel) at (4.375, -3) {PME};

% Draw factor nodes
\node[draw=none] (fftlabel) at (-3.125, 1.7) {FFT-Based};
\node[draw=none] (fftlabel) at (-3.125, 1.3) {Factors};
\node[check](c1) at (-1.75, 1.5) {$a_1$};
\node[check](c2) at (0, 1.5) {$a_2$};
\node[check](c3) at (1.75, 1.5){$a_3$};

% Draw section nodes
\node[section](s1) at (-2.75, -0.5) {$s_1$};
\node[draw=none](dots) at (-1.375, -0.5) {$\ldots$};
\node[section](s_ell) at (0, -0.5) {$s_\ell$};
\node[draw=none](dots2) at (1.375, -0.5){$\ldots$};
\node[section](s3) at (2.75, -0.5) {$s_{L}$};

% Draw graph edges
\draw (c1.south) -- (s1.north);
\draw (c1.south) -- (s_ell.north);
\draw (c2.south) -- (s1.north);
\draw (c2.south) -- (s_ell.north);
\draw (c2.south) -- (s3.north);
\draw (c3.south) -- (s1.north);
\draw (c3.south) -- (s_ell.north);
\draw (c3.south) -- (s3.north);

% Draw graph edges
\draw[dashed] (c1.south) -- (-2,0.5);
\draw[dashed] (c1.south) -- (-1.75,0.5);
\draw[dashed] (c1.south) -- (-1.5,0.5);
\draw[dashed] (c1.south) -- (-1.25,0.5);
\draw[dashed] (c2.south) -- (-0.85,0.5);
\draw[dashed] (c2.south) -- (-0.55,0.5);
\draw[dashed] (c2.south) -- (-0.25,0.5);
% \draw[dashed] (c2.south) -- (0,0.5);
\draw[dashed] (c2.south) -- (0.25,0.5);
\draw[dashed] (c2.south) -- (0.55,0.5);
\draw[dashed] (c2.south) -- (0.85,0.5);
\draw[dashed] (c3.south) -- (1.25,0.5);
\draw[dashed] (c3.south) -- (1.5,0.5);
\draw[dashed] (c3.south) -- (1.75,0.5);
\draw[dashed] (c3.south) -- (2,0.5);

% Draw PME nodes
\node[quantity] (pme1label1) at (-2.55, -2.9) {};
\node[quantity] (pme1label2) at (-2.65, -3.0) {};
\node[quantity] (pme1label3) at (-2.75, -3.1) {$\hat{s}_g \left(\; \cdot \;, \; \cdot \;, \tau_t \right)$};

\node[quantity] (pme2label1) at (0.2, -2.9) {};
\node[quantity] (pme2label2) at (0.1, -3.0) {};
\node[quantity] (pme2label3) at (0, -3.1) {$\hat{s}_g \left(\; \cdot \;, \; \cdot \;, \tau_t \right)$};

\node[quantity] (pme3label1) at (2.95, -2.9) {};
\node[quantity] (pme3label2) at (2.85, -3.0) {};
\node[quantity] (pme3label3) at (2.75, -3.1) {$\hat{s}_g \left(\; \cdot \;, \; \cdot \;, \tau_t \right)$};

% Draw PME to section node edges
\draw[densely dashed] (s1) -- (pme1label1.north);
\draw[densely dashed] (s1) -- (pme1label2.north);
\draw[densely dashed] (s1) -- (pme1label3.north);
\draw[densely dashed] (s_ell) -- (pme2label1.north);
\draw[densely dashed] (s_ell) -- (pme2label2.north);
\draw[densely dashed] (s_ell) -- (pme2label3.north);
\draw[densely dashed] (s3) -- (pme3label1.north);
\draw[densely dashed] (s3) -- (pme3label2.north);
\draw[densely dashed] (s3) -- (pme3label3.north);

% Draw labels to various graph edges
\draw [->] (-2.85, 0) -- (-2.1, 1);
\node[draw=none, rotate=53] (mu_s_a) at (-2.65, 0.65) {$\muv_{s_1 \rightarrow a_1}$};
\draw [<-] (2.85, 0) -- (2.1, 1);
\node[draw=none, rotate=-53] (mu_a_s) at (2.65, 0.65) {$\muv_{a_3 \rightarrow s_{L}}$};

\draw[->] (-2.4, -1.25) -- node[right] {$\muv_{s_1}$} (-2.4, -2);
\draw[<-] (-2.95, -1.25) -- node[left] {$\lambdav_1$} (-2.95, -2);

\draw[->] (0.35, -1.25) -- node[right] {$\muv_{s_{\ell}}$} (0.35, -2);
\draw[<-] (-0.2, -1.25) -- node[left] {$\lambdav_{\ell}$} (-0.2, -2);

\draw[->] (3.1, -1.25) -- node[right] {$\muv_{s_L}$} (3.1, -2);
\draw[<-] (2.55, -1.25) -- node[left] {$\lambdav_L$} (2.55, -2);

% Draw inputs/outputs to each set of PMEs
\draw[<-,rounded corners] (-2.875, -3.625) -- (-2.875, -4.625) -- node[above,xshift=-1.5mm] {$\rv_g^{(t)}(1)$} (-3.75, -4.625);
\draw[->,rounded corners] (-2.625, -3.625) -- (-2.625, -4.75) -- node[below,xshift=-3.5mm] {$\sv_g^{(t+1)}(1)$} (-1.75, -4.75) -- (-1.75, -5.25);

\draw[<-,rounded corners] (-0.125, -3.625) -- (-0.125, -4.625) -- node[above,xshift=-1.5mm] {$\rv_g^{(t)}(\ell)$} (-1, -4.625);
\draw[->,rounded corners] (0.125, -3.625) -- (0.125, -4.75) -- node[below,xshift=-3.5mm] {$\sv_g^{(t+1)}(\ell)$} (1, -4.75) -- (1, -5.25);

\draw[<-,rounded corners] (2.625, -3.625) -- (2.625, -4.625) -- node[above,xshift=-2mm] {$\rv_g^{(t)}(L)$} (1.75, -4.625);
\draw[->,rounded corners] (2.875, -3.625) -- (2.875, -4.75) -- node[below,xshift=-4mm] {$\sv_g^{(t+1)}(L)$} (3.75, -4.75) -- (3.75, -5.25);

% Draw parallel-serial box 
% \node[psbox] (ps) at (0.1, -3.75) {Serial $\longleftrightarrow$ Parallel};

% Draw labels for PME/FG interactions
% \draw[->] (-0.1, -5.0) -- (-0.1, -4.25);
% \node[draw=none] (inputlabel) at (-0.50, -4.625) {$\mathbf{r}_g^{(t)}$};
% \draw[<-] (0.3, -5.0) -- (0.3, -4.25);
% \node[draw=none] (outputlabel) at (1.25, -4.625) {$\boldsymbol{\eta}_g^{(t)}\left(\mathbf{r}_g^{(t)}\right)$};
% \draw[->] (-0.2, -5.0) -- (-0.2, -4.25);
% \node[draw=none] (inputlabel) at (-0.60, -4.625) {$\mathbf{r}_g^{(t)}$};
% \draw[<-] (0.2, -5.0) -- (0.2, -4.25);
% \node[draw=none] (outputlabel) at (1.15, -4.625) {$\boldsymbol{\eta}_g^{(t)}\left(\mathbf{r}_g^{(t)}\right)$};

% JF Suggestions
% \draw[->] (s3.south) -- (2.5, -2.65);
% \draw (s3.south) -- (2.7, -2.25);
% \draw (s3.south) -- node[right] {$\qv_L$} (2.8, -2.15);
% \draw[->, rounded corners] (1.7, -4) -- (1.7, -3) -- node[left] {$\rv_L$} (1.7, -2) -- (s3);
% \draw[->, rounded corners] (1.7, -3.5) -- (2.85, -3.5) -- (2.85, -2.9);
% \draw[rounded corners] (1.7, -3.5) -- (2.95, -3.5) -- (2.95, -3.25);
% \draw[rounded corners] (1.7, -3.5) -- (3.05, -3.5) -- (3.05, -3.25);
%(2.85, -2.5);
% \draw[->] (2.85, -2) -- node[right] {$\rv_L$} (s3.south);
% Add input/output labels

% \draw[->] (-2.95, -4.5) -- (-2.95, -3.25);
% \node[draw=none] (rlabel) at (-3.25, -4) {$\rv_1^{(t)}$};

% \draw[<-] (-2.55, -4.5) -- (-2.55, -3.25);
% \node[draw=none] (etalabel) at (-1.85, -4) {$\boldsymbol{\eta}\left(\rv_1^{(t)}\right)$};

\end{tikzpicture}
    \caption{
        This figure illustrates the denoiser employed for group $g$ within the AMP iterate for coded demixing.
        The PME of \eqref{eq:OriginalPME} is applied to the effective observation $\rv$ using uninformative priors to obtain local observations $\{\lambdav_\ell : \ell \in [L]\}$, which initialize the outer LDPC factor graph. 
        After running belief propagation, messages $\muv_{s_\ell}$ are sent from the variable nodes in the factor graph to the PME for each section; these messages are used to generate informative priors. 
        Equipped with more accurate priors, the PME is run on the effective observation $\rv$ once more, yielding $\etav\left( \rv \right)$.
    }
    \label{fig:BPonFactorGraph}
\end{figure}
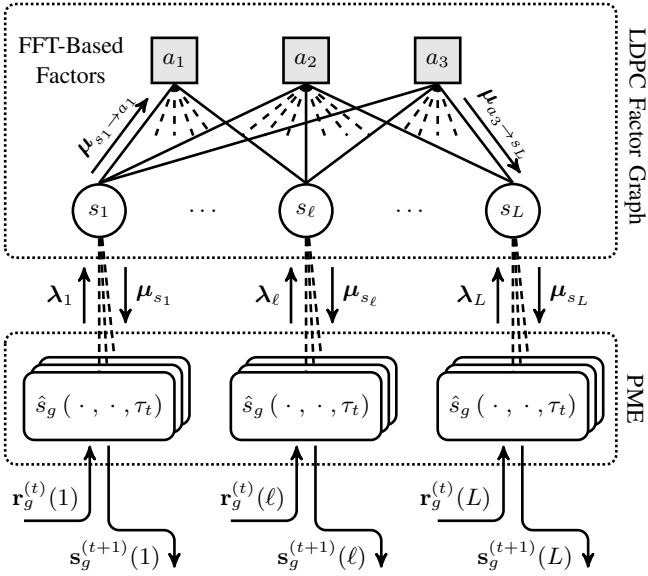

In \cite{amalladinne2020unsourced}, Amalladinne et al.\ show that, when there is only one group, the structure of the outer LDPC code can be exploited to improve the PME denoiser.
Specifically, one round of BP may be run on a suitably designed outer factor graph to generate more accurate priors.
This helps AMP converge to a parity-consistent solution, and it has been shown to improve the performance of CCS by $1$~dB when $K = 100$. 
We wish to extend these developments to coded demixing by generating priors for each group through message-passing on each group's outer LDPC factor graph.

We only briefly summarize the message passing rules associated with the LDPC outer codes because the details of message passing on factor graphs are well established (e.g., \cite{kschischang2001factorgraph}).
We also omit the group subscript $g$ for this part to lighten notation.
While every group has its own LDPC outer code, belief propagation acts in a similar fashion within each group, with no explicit interactions between groups.
The fact that the LDPC factor graphs are disjoint across groups should prevent any confusion about group indexing.
Variable nodes are initialized with local estimates $\lambdav_{\ell}\left(k\right) = \hat{s}_{g}(q_g, \rv_g(\ell, k), \tau_t)$.
One round of belief propagation is then performed with check to variable messages given by
\begin{equation} \label{eq:BP-Check2Variable}
    \muv_{a \to s_{\ell}} (k)
    = \sum_{\kv_{a}: k_{\ell} = k} \Psi_{a} \left( \kv_{a} \right)
    \prod_{s_j \in N(a) \setminus s} \muv_{s_j \to a} (k_j),
\end{equation}
where $\Psi_{a}$ is an indicator function that enforces parity consistency, $k$ is an index within a section, and $\kv_{a}$ is a collection of section indices for the neighbors $N(a)$ of check node $a$.
Messages from variable nodes to check nodes are given by
\begin{equation} \label{eq:BP-Variable2Check}
    \muv_{s_{\ell} \rightarrow a} (k)
    \propto \lambdav_{\ell} (k) \prod_{a_p \in N(s_{\ell}) \setminus a} \muv_{a_p \to s_{\ell}} (k),
\end{equation}
where the `$\propto$' indicates a normalization of the message.
After running one round of BP, the belief associated with section $\ell$ may be obtained as follows
\begin{equation} \label{eq:EquivalentPriors}
    \muv_{s_{\ell}}(k) = \prod_{a \in N(s_{\ell})} \muv_{a \to s_{\ell}} (k) .
\end{equation}
Once normalized, \eqref{eq:EquivalentPriors} may be used as a more informative prior $q$ in the PME of \eqref{eq:OriginalPME}.
Explicitly,
\begin{equation} \label{equation:BP-priors}
    \qv_g(\ell, k) = 1 - \left(1 - \frac{\muv_{s_\ell}\left(k\right)}{\| \muv_{s_\ell} \|_1}\right)^{K_g} .
\end{equation}
Note that log-domain Fourier transform decoding \cite{davey1998low,song2003reduced} is employed for efficient computation of BP. 
Accordingly FFT-based factors are employed in the bipartite graph representing the LDPC code.
\noindent
An illustration of message passing on a generic bipartite graph is provided in Fig.~\ref{fig:BPonFactorGraph}. 

Combining the PME of \eqref{eq:OriginalPME} with the priors computed dynamically using BP results in the smooth denoiser found in \cite[Definition 6]{amalladinne2020unsourced}.
Accounting for the various sparsity domains in coded demixing and their distinct outer LDPC codes, we get group-specific non-separable denoisers of the form
\begin{equation}
    \label{eq:dynamicdenoiser}
    \etav^{(t)}_{g}(\rv_g) = \hat{\sv}_{g}(1, \rv_g, \tau_t)\cdots\hat{\sv}_{g}(L_g, \rv_g, \tau_t),
    % \etav^{(t)}_{g}(\rv_g) = \hat{\sv}_{g}(\ell, \rv_g, \tau_t) % \cdots\hat{\sv}_{g}(\rv_g(L_g), \tau_t)
\end{equation}
where individual components are based on \eqref{eq:OriginalPME} with
\begin{equation*}
    \begin{split}
        &\hat{\sv}_{g}(\ell, \rv_g, \tau_t) = \left( \hat{s}_{g}(\qv_g(\ell, k), \rv_g(\ell, k), \tau_t) : k \in [2^{v_g}] \right) % 0, \ldots, 2^{v_g} - 1 \right)
    \end{split}
\end{equation*}
and priors are obtained from \eqref{equation:BP-priors} with one round of message passing on the LDPC factor graph.
This same approach is used for all groups $\{g \in [G]\}$, though each group runs BP on its own factor graph and power levels may vary across groups.
This explains why the subscript $g$ must be included in the definition of $\etav^{(t)}_g(\rv)$. 
It has been shown that this denoiser is Lipschitz continuous; thus, it is well-suited for AMP. 
In computing the AMP residual in \eqref{eq:ampresidual}, the Onsager correction term requires the divergence of the denoiser, which is provided below,
\begin{equation}
    \label{eq:denoiserdivergence}
    \mathrm{div}\Dm\etav^{(t)} = \frac{1}{\tau_t^2}\left(
    \| \Dm^2\etav^{(t)}(\rv) \|_1 - \| \Dm\etav^{(t)}(\rv) \|_2^2 \right) .
\end{equation}
Note that the form of the Onsager term remains remarkably simple despite the presence of several groups.
This stems from the additive structure of both $\| \cdot \|_1$ and $\| \cdot \|_2^2$, which yields
\begin{equation*}
    \sum_{g \in [G]} \mathrm{div} \left( d_g \etav_g^{(t)} \right)
    = \mathrm{div}\Dm\etav^{(t)} .
\end{equation*}
In this sense, the compact notation of \eqref{eq:denoiserdivergence} is due, not only to the demixing formulation, but also to the specifics of our denoiser.
It is noted that \eqref{eq:denoiserdivergence} holds only when the number of BP iterations per AMP step is strictly less than the length of the shortest cycle on the outer graph. 
The derivation of \eqref{eq:denoiserdivergence} is provided in Appendix \ref{appendix:onsagerderivation}. 

After running several iterations of the AMP algorithm via \eqref{eq:ampresidual} and \eqref{eq:ampstateupdate}, the decoder produces estimates for $\{\sv_g : g\in [G]\}$.
To get the transmitted messages, the receiver must then extract valid codewords by finding parity-consistent indices across the $L_g$ recovered sections of $\sv_g$.
The indices associated with the $K_g + \delta$ largest values in the first section of $\sv_g$ are identified and the following actions are repeated for each of the selected indices. 
The root section of the LDPC factor graph is initialized with the standard basis vector $\ev_i$ where $i$ is the selected index and the other $L_g - 1$ sections are initialized with the output of AMP.
Several BP iterations on the graph of the outer code are then performed, which yields candidate sections.
If a parity-consistent codeword is obtained, that codeword is added to the list of recovered codewords along with its associated likelihood. 
Once this process has been performed for all $G$ bins, the recovered codewords from all bins are collected and ordered in terms of decreasing likelihood.
Finally, the top $K$ codewords are retained as $\hat{\mathcal{W}}(\yv)$. 

\begin{remark}
\label{remark:hadamardmatrices}
While the theory of AMP and state evolution applies to random Gaussian matrices,
it is common in practice to generate each $\Am_g$ by randomly sampling the rows of a $2^{v_g} \times 2^{v_g}$ Hadamard matrix (excluding the row of all ones), instead of generating the sensing matrices $\{\Am_g : g \in [G]\}$ with random Gaussian entries.
This enables the use of Fast Walsh-Hadamard Transform techniques within the AMP iterations \eqref{eq:ampresidual} and \eqref{eq:ampstateupdate}, which decreases both the computational complexity and the required memory.
The fact that AMP works well for certain random matrices with non-Gaussian entries was shown empirically in \cite{donoho2009message}.
\end{remark}

Having established a framework for coded demixing, we proceed with the description of two application scenarios: accommodating multiple classes of users within a network, and improving the PUPE performance of a single-class network via stochastic binning of users.  
We begin with an examination of the multi-class URA network. 

\section{Multi-Class Networks via Coded Demixing}
\label{section:MulticlassviaCodedDemixing}

In its original definition, the URA model is composed of $K_{\mathrm{tot}}$ \textit{homogeneous} users.
That is, all $K_{\mathrm{tot}}$ users have the exact same power budget, message length, and code requirements. 
An interesting and highly applicable twist to the original URA setting is that of a \textit{heterogeneous} URA model (HetURA), as described in \cite{hao2020exploration}, where devices are partitioned into classes with varying power, data, and code requirements.

These classes may either be completely independent or connected together in some fashion. 
As an example of independent classes, consider an industrial setup where a central base station receives sensor data from various sections of a warehouse, and each section is located at a different distance from the base station.
To offset the free space path loss incurred by the additional distance, sensors in further away sections employ a lower code rate when communicating with the base station. 
In this example, each class consists of the group of sensors that employ a specific code rate.

As an example of connected classes, consider the task of sending a long URA message over the channel, say on the order of $256$ bits.
Unfortunately, off-the-shelf CS is not able to handle the dimensionality of this problem.
One option to circumvent this issue is to apply the technique presented in \cite{ebert2020hybrid}; another option is to employ multiple, concatenated classes. 
Specifically, let the transmitter divide its message into sections and link each section together with an additional outer code. 
Then, let the transmitter send the first part of its message as part of class one, the second as part of class two, and so on until the final portion of its message is sent via the class $\varrho$.
After recovering all the received messages, the decoder may use side information about classes $1, \ldots, \varrho$ and the outermost linking parities to properly stitch the very large message together.

For ease of exposition, we study only the case of independent classes in our simulations, but the results presented in this article may be extended to the connected class case with minimal modifications.

\subsection{Prior Work}

After introducing the HetURA model in \cite{hao2020exploration}, Hao et al.\ propose a scheme for the case when two classes of users exist with different power levels. 
Therein, the high-energy users employ a two-layer superposition modulation and the low-energy users employ the same base constellation as the high-energy users.
The top level of the superposition constellation is decoded, followed by the lower level with the help of SIC.
Finally, a CCS-style tree code is employed as an outer code to disambiguate messages. 
A version of this problem is also studied by Huang et al.\ in \cite{huang2021iterative}, wherein each user transmits its message $L$ times in random slots.  
At the receiver, power-domain NOMA techniques are leveraged to handle collisions within a single slot and inter-slot SIC is applied to improve performance.  
In \cite{amalladinne2020multiclass}, this problem is considered from a coded demixing perspective. 
It is upon the work of Amalladinne et al.~\cite{amalladinne2020multiclass} that we wish to expound in this article. 

\subsection{Proposed Approach}

The coded demixing model presented in Section~\ref{section:SystemModel} may be applied to the HetURA channel in the following manner.  
Let the $K_{\mathrm{tot}}$ total users be clustered into $G$ groups, where each group consists of devices from the same class.
We note that a class of users can be broken up into multiple groups, but a single group cannot have devices from multiple classes. 
Thus, each group $g \in [G]$ has an associated amplitude level $d_g$, message length $w_g$, coding rate $R_g$, codeword length $v_g$, and sensing matrix $\Am_g$.
Furthermore, to conform to a widely adopted convention, it is assumed that the number of active users per group is known to the receiver.

\subsection{Two-Class System Model}

For illustrative purposes, we consider a scheme with two groups, each containing $25$ active users.
Group~1 features a message length of $128$ bits and a coding rate of $1/2$, and group~2 features a message length of $96$ bits with a coding rate of $3/8$. 
The total number of sections (parity + information) within each graph is $16$, and the length of each section is $2^{16}$. 
The power budget $P$ is obtained via the relationship $\frac{E_{\mathrm{b}}}{N_0} = \frac{nP}{2w_g}$. 
The transmitter outer/inner encoding operations proceed exactly as outlined in Section \ref{section:SystemModel}.
At the receiver, AMP decoding and message disambiguation proceed as outlined above, with the exception that codewords recovered from group 1 are kept separate from codewords recovered from group 2. 
The PUPE is computed as in \eqref{eq:pupe} for each of the two groups individually.  

\begin{figure}[t]
    \centering
    \begin{tikzpicture}
\definecolor{colorTIN2}{RGB}{128,0,0}
\definecolor{colorTIN1}{RGB}{192,0,0}
\definecolor{colorSIC2}{RGB}{0,128,0}
\definecolor{colorSIC1}{RGB}{0,192,0}
\definecolor{colorJD2}{RGB}{0,0,128}
\definecolor{colorJD1}{RGB}{0,0,192}
\begin{semilogyaxis}[
font=\small,
width=7cm,
height=5.5cm,
scale only axis,
every outer x axis line/.append style={white!15!black},
every x tick label/.append style={font=\color{white!15!black}},
xmin=1.8,
xmax=2.8,
xtick={1.8,2,2.2,2.4,...,2.8},
xlabel={$E_{\mathbf{b}}/N_0$ (dB)},
xmajorgrids,
xminorgrids,
every outer y axis line/.append style={white!15!black},
every y tick label/.append style={font=\color{white!15!black}},
ymax=1,
ytick={0.01, 0.1, 1.0},
ylabel={Per-User Error Rate $P_{\mathrm{e}}$},
ymajorgrids,
yminorgrids,
legend style={at={(1, 1)},anchor=north east,draw=black, fill=white, legend cell align=left,font=\scriptsize}
]
\addplot [color=colorTIN1, dotted,line width=2.0pt,mark size=1.4pt,mark=square,mark options={solid}]
  table[row sep=crcr]{
  1.8 0.352\\
  2 0.2972\\
  2.2 0.2364\\
  2.4 0.1856\\
  2.6 0.1496\\
  2.8 0.1252\\
};
\addlegendentry{Group 1: TIN};

\addplot [color=colorTIN2,densely dotted,line width=2.0pt,mark size=1.4pt,mark=star,mark options={solid}]
  table[row sep=crcr]{
  1.8 0.3416\\
  2 0.3024\\
  2.2 0.2604\\
  2.4 0.2072\\
  2.6 0.1856\\
  2.8 0.1376\\
};
\addlegendentry{Group 2: TIN};

\addplot [color=colorSIC1,dashdotted,line width=2.0pt,mark size=1.4pt,mark=o,mark options={solid}]
  table[row sep=crcr]{
  1.8 0.3392\\
  2 0.28\\
  2.2 0.2384\\
  2.4 0.1968\\
  2.6 0.1568\\
  2.8 0.1244\\
};
\addlegendentry{Group 1: SIC};

\addplot [color=colorSIC2,densely dashdotted,line width=2.0pt,mark size=1.4pt,mark=triangle,mark options={solid}]
  table[row sep=crcr]{
  1.8 0.2828\\
  2 0.2388\\
  2.2 0.1864\\
  2.4 0.148\\
  2.6 0.108\\
  2.8 0.0848\\
};
\addlegendentry{Group 2: SIC};

\addplot [color=colorJD1,densely dashed,line width=2.0pt,mark size=1.4pt,mark=diamond,mark options={solid}]
  table[row sep=crcr]{
  1.8 0.2024\\
  2 0.1488\\
  2.2 0.1092\\
  2.4 0.0888\\
  2.6 0.06\\
  2.8 0.0492\\
};
\addlegendentry{Group 1: CD};

\addplot [color=colorJD2,solid,line width=2.0pt,mark size=1.4pt,mark=pentagon,mark options={solid}]
  table[row sep=crcr]{
  1.8 0.1344\\
  2 0.092\\
  2.2 0.0712\\
  2.4 0.0464\\
  2.6 0.038\\
  2.8 0.0376\\
};
\addlegendentry{Group 2: CD};

% \addplot [color=colorTIN1, dashdotted,line width=2.0pt,mark size=1.4pt,mark=square,mark options={solid}]
%   table[row sep=crcr]{
%   1.8 0.3468\\ % 
%   2 0.2998\\%
%   2.2 0.2484\\%
%   2.4 0.1964\\%
%   2.6 0.1676\\%
%   2.8 0.1314\\%
% };
% \addlegendentry{TIN};

% \addplot [color=colorSIC2,densely dashdotted,line width=2.0pt,mark size=1.4pt,mark=triangle,mark options={solid}]
%   table[row sep=crcr]{
%   1.8 0.311\\
%   2 0.2594\\
%   2.2 0.2124\\
%   2.4 0.1724\\
%   2.6 0.1324\\
%   2.8 0.1046\\
% };
% \addlegendentry{SIC};

% \addplot [color=colorJD2,solid,line width=2.0pt,mark size=1.4pt,mark=pentagon,mark options={solid}]
%   table[row sep=crcr]{
%   1.8 0.1684\\
%   2 0.1204\\
%   2.2 0.0902\\
%   2.4 0.0676\\
%   2.6 0.049\\
%   2.8 0.0434\\
% };
% \addlegendentry{Coded Demixing};
\end{semilogyaxis}
\end{tikzpicture}
    \vspace{-3mm}
    \caption{
    This plot compares the PUPE of various multi-class URA schemes.
    Two classes are considered, each with $25$ active devices.
    In group~1, $w_1 = 128$~bits and $R_1 = 1/2$.  
    In group~2, $w_2 = 96$~bits and $R_2 = 3/8$.
    Coded demixing outperforms other known communication schemes.}
    \label{fig:MultiClassResults}
\end{figure}
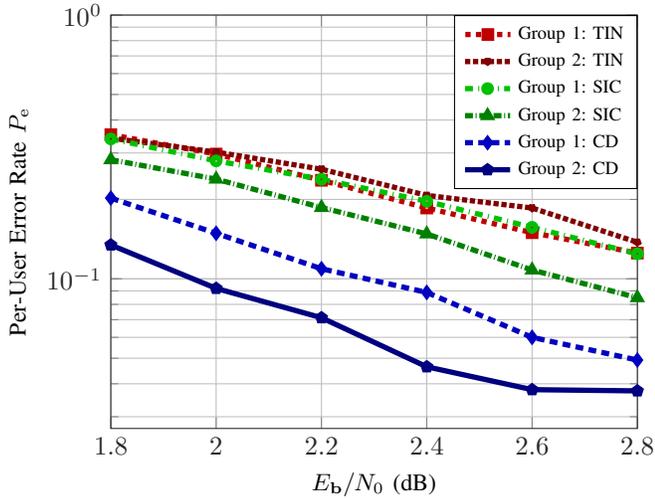

\subsection{Two-Class Simulation Results}

We compare the performance of multi-class CCS via coded demixing to other existing communication schemes, including treating interference as noise (TIN) and single-class decoders with SIC. 
In TIN, the decoder for group~1 operates on received vector $\yv$, and views the contribution of group~2 as additional noise.
Similarly, the decoder for group~2 operates on $\yv$ and treats interference from group~1 as noise. 
In SIC, the decoder first produces an estimate $\hat{\sv}_1$ of the signal sent by group~$1$ based on observation $\yv$.
It then subtracts the estimated contribution of group~1 from $\yv$, yielding the residual $\Tilde{\yv} = \yv - \Phi_1\hat{\sv}_1$.
The decoder then produces an estimate $\hat{\sv}_2$ of the signal transmitted by group~2 from residual $\tilde{\yv}$. 

Fig.~\ref{fig:MultiClassResults} shows that coded demixing outperforms both SIC and TIN in both groups by up to $0.6$~dB.
In addition to providing a striking PUPE improvement, the complexity of the coded demixing solution is very comparable to the complexity of SIC or TIN.
Altogether, a judicious use of coded demixing forms a promising paradigm for multi-class URA, and it currently offers state-of-the-art performance.

\section{Enhanced Single-Class CCS via Coded Demixing and Stochastic Binning}
\label{section:StochasticBinningandCodedDemixing}

In Section~\ref{section:MulticlassviaCodedDemixing}, the coded demixing framework was considered as a means to enable multiple classes of devices to simultaneously use the same network resources. 
In this section, we demonstrate how the coded demixing framework may be used to enhance the PUPE performance of a network with a single class of users by randomly partitioning active devices into groups.

Within CCS, $K$ codewords are present on the outer LDPC factor graph at once during message passing \cite{amalladinne2020unsourced}. 
This results in a form of mixing that makes it difficult for belief propagation to distinguish between codewords and degrades performance.
Though the belief propagation algorithm may be modified slightly to facilitate multi-user decoding, reducing the number of simultaneous codewords on the factor graph seems to constitute a better way to improve the performance of the outer decoder. 
As a means to accomplish this goal, the collection of active devices may stochastically partition themselves into groups, and subsequently adopt the coded demixing framework introduced above (see also \cite{ebert2021stochastic}). 
We elaborate on this option below.

\subsection{Stochastic Binning}

We propose a scheme in which there are $G$ homogenous groups with separate sensing matrices and outer factor graphs. 
We stress that these groups share the same transmit power, message length, number of sections, and coding rate; that is, $d_g = d_h$, $w_g = w_h$, $v_g = v_h$, $L_g = L_h$, $R_g = R_h$ for all $g, h \in [G]$. 
To emphasize the fact that the groups are homogeneous, we employ the term \emph{bin} instead of \emph{group}. 

\begin{figure}[ht!]
    \centering
    \begin{tikzpicture}
    [
    font=\small, >=stealth', line width=1pt, draw=black,
    block/.style={rectangle, draw, minimum height=5.5mm, minimum width=7mm, rounded corners},
    section/.style={circle, minimum size=7mm, draw=black},
    check/.style={rectangle, minimum height=6mm, minimum width=6mm, draw=black, fill=gray!20, rounded corners},
    trivialcheck/.style={rectangle, minimum height=4mm, minimum width=4mm, draw=black, fill=gray!20},
    section/.style={circle, minimum size=7mm, draw=black},
    ]

% Draw fancy boxes
\draw[draw=black, fill=white, densely dotted, rounded corners] (1.0, 4.0) rectangle (6.7, 6);
\node[draw=none, rotate=90] (binidlabel) at (0.75, 5) {Bin ID};
\draw[draw=black, fill=white, densely dotted, rounded corners] (1.0, -0.125) rectangle (6.875, 3.75);
\node[draw=none, rotate=90] (fig1subblocklabel) at (0.75, 1.8125) {Fig.~\ref{figure:EncodingProcess} Encoding};

% Stage 1: user's starting message block
\node[block](wj) at (0.5, 3.375) {$\mathbf{w}_j$};

% Stage 2: user's message fragments
\node[block, rotate=90](wjbin) at (1.5, 5.0) {$\mathbf{w}_j\left[1\mathpunct{:}w_0\right]$};
\draw[draw=black, fill=white, rounded corners] (1.25, 0) rectangle (1.75, 3.5);
\node[draw=none, rotate=90](wjmainlabel) at (1.5, 1.75) {$\mathbf{w}_j\left[w_0+1\mathpunct{:}w\right]$};

% Stage 1-2 connections
\draw[->,shorten >=0.5cm] (wj.east) -- (1.25, 5.0);
\draw[->,shorten >=0.5cm] (wj.east) -- (1.25, 2.0);

% Stage 3
\draw[draw=black, fill=white, rounded corners] (2.25, 0) rectangle (4.25, 3.5);
\node[draw=none, rotate=90] (ldpcenclabel) at (3.25, 1.75) {LDPC Encoder};

% % Stage 2-3 connections
\draw[->] (1.75, 3) -- (2.25, 3);
\draw[->] (1.75, 2.5) -- (2.25, 2.5);
\node[draw=none] (vdots231) at (2.0, 1.5) {$\vdots$};
\draw[->] (1.75, 0.5) -- (2.25, 0.5);
\draw[->, rounded corners] (wjbin.south) -- (3.25, 5.0) -- (3.25, 3.5);

% Stage 4
\node[block] (eg) at (5.75, 5.0) {$\mathbf{e}_g$};
\node[draw=none] (flabel) at (3.625, 5.25) {$f\left(\cdot\right)$};
\draw[draw=black, fill=white, rounded corners] (4.75, 0) rectangle (6.75, 3.5);
\node[draw=none, rotate=90] (csenclabel) at (5.75, 1.75) {CS Encoder};

% Stage 3-4 connections
\draw[->] (wjbin.south) -- (eg.west);
\draw[->] (4.25, 3) -- (4.75, 3);
\draw[->] (4.25, 2.5) -- (4.75, 2.5);
\node[draw=none] (vdots231) at (4.5, 1.5) {$\vdots$};
\draw[->] (4.25, 0.5) -- (4.75, 0.5);
\draw[->, rounded corners] (wjbin.south) -- (4.25, 5.0) -- (5.75, 4.25) -- (5.75, 3.5);
\node[draw=none] (aglabel) at (6.0, 4.25) {$\mathbf{A}_g$};

% Stage 5 
\node[block, rotate=90] (concat) at (7.5, 3.375) {concat};

% stage 4-5 connections
\draw[->, rounded corners] (eg.east) -- (7.5, 5.0) -- (concat.east);
\draw[->, rounded corners] (6.75, 1.75) -- (7.5, 1.75) -- (concat.west);

% Stage 6
\node[draw, circle] (MAC) at (8.5, 3.375) {$\Sigma$};

% Stage 5-6 connections
\draw[->] (concat.south) -- (MAC.west);
\draw[->] (8.5, 4.375) -- (MAC.north);
\node[draw=none] (awgn) at (8.5, 4.5) {$\mathbf{z}$};
\draw[->, rounded corners] (8.0, 2.2375) -- (8.0, 2.75) -- (MAC.south west);
\draw[->, rounded corners] (8.0, 4.375) -- (8.0, 4) --  (MAC.north west);
\draw[->] (MAC.south) -- (8.5, 1.5);
\node[draw=none] (ylabel) at (8.5, 1.25) {$\mathbf{y}$};

\end{tikzpicture}
    \caption{
        This figure illustrates the encoding process for coded demixing with stochastic binning. 
        Each user's message is separated into two parts: the first $w_0$ bits are used to select a bin and the remaining $\left(w - w_0\right)$ bits are encoded as described in Fig.~\ref{figure:EncodingProcess}.
        We note that the choice of bin determines the outer LDPC code employed as well as the sensing matrix $\Am_g$ used for CS encoding.
        Because the receiver does not know how many users have selected each bin a priori, a bin identification sequence must be sent so the distribution of active users across bins can be estimated. 
    }
    \label{fig:StochasticBinningEncoding}
\end{figure}
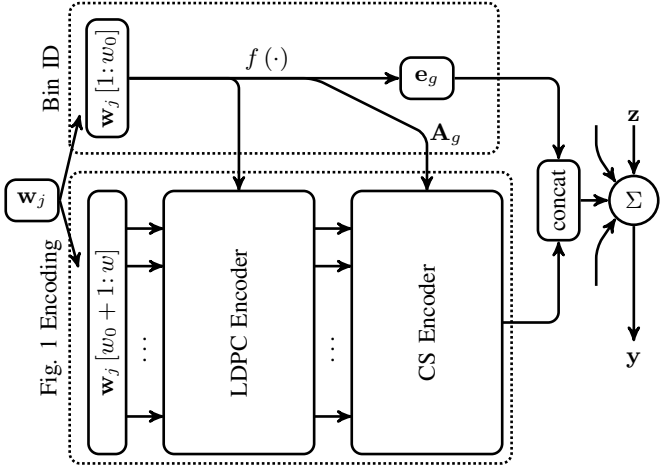

After the $K$ active users have generated their messages, every user selects a bin based on the first $w_0 = \log_2(G)$ bits of its information message.
We restrict our treatment of this approach to values of $G$ that are powers of two;
other values of $G$ are possible, but would require slight adjustments to the proposed algorithm.
The distribution of users across bins $\{K_g : g \in [G]\}$ assumes a multinomial distribution with $K$ trials and probabilities $p_g = 1/G$ for all $g \in [G]$. % $p_1 = \cdots = p_G = \frac{1}{G}$. 
Fig.~\ref{fig:StochasticBinningEncoding} depicts the bin selection and encoding processes.

We stress that the process of assigning active users to bins must be performed stochastically since
there is no a-priori  knowledge on what users are currently active. 

Although within the URA common task framework the total number of active users $K$ is known to the receiver, the number of users per bin $\{K_g : g \in [G]\}$ is unknown.
Rather, this information must be inferred at the destination. 
To enable accurate bin occupancy estimation, every active user simultaneously transmits a length-$G$ standard basis vector $\ev_g$, where the index of the single unitary entry in $\ev_g$ corresponds to the bin that the user has selected.
Within the URA framework, the expected total power used per device is upper-bounded by $nP$.
Consequently, the power devoted to transmitting the bin identification sequence must be subtracted from the power allocated to sending the information message.
Numerical simulations indicate that allocating $0.2\,\%$ of the total power to the bin occupancy estimation task allows for good estimates $\{\hat{K}_g : g \in [G]\}$ without significantly degrading the PUPE performance. 
At the receiver, a standard linear minimum mean square error (LMMSE) estimator is used to jointly estimate the number of users per bin $\{\hat{K}_g : g \in [G]\}$. 
These estimates are used throughout the AMP decoding algorithm, as presented in Section~\ref{section:SystemModel}. 

After transmitting their bin identification sequences, the set of active users transmit their encoded messages over the GMAC. 
After estimating the number of users per bin, the AMP-based coded demixing algorithm of Section.~\ref{section:CodedDemixing} is employed and $\hat{K}_g + \delta$ potential messages are retained within bin~$g$.
At this stage, we point out that $w_0$ bits of information are conveyed per message via the choice of a bin; simply stated, every user within bin~$g$ must have the same first $w_0$ bits, which must equal the binary representation of $g$.
The bin identity can therefore be leveraged as an additional error-detecting code and the list of candidate messages within each bin may be pruned by removing all messages whose first $w_0$ bits do not correspond to $g$.
After removing parity-inconsistent messages within each bin, the $G$ lists of candidate messages, one for each bin, are merged into a single list of recovered messages.
The aggregate list is sorted in terms of decreasing likelihood, and the top $K$ messages from this sorted list are taken as the $K$ reported messages by the base station.
To further improve performance, one outer loop of SIC may be employed, wherein approximately $70\,\%$ of the recovered messages are retained during the first round of SIC.  
After decoding these messages, their contributions are subtracted from the received vector $\yv$, and the decoding algorithm is run one more time to recover the remaining $30\,\%$ of the messages. 

\begin{remark}
\label{remark:binlimits}
As will be shown shortly, in some regimes, increasing the number of bins tends to improve performance. 
However, one must tread carefully because AMP is known to only converge within certain regions in terms of undersampling and sparsity ratios~\cite{donoho2009message}.
This depends on the problem formulation and the denoiser, yet traditional settings offer a cautionary tale as to what can be accomplished.
For our scheme, the sparsity is equal to $\sum_{g \in [G]}K_gL/n = KL/n$ and the undersampling ratio is given by $n/GL2^v$.  
As the number bins $G$ is increased, the sparsity level remains constant but the undersampling ratio goes to zero.
This can affect the ability of AMP to handle noise and, at worst, can cause the operating point of the system to fall out of the convergence region.

Furthermore, low cross-coherence among $\{\Am_g : g \in [G]\}$ is a critical underlying assumption of the theory of demixing and, as $G$ increases, the cross-coherence among $\{\Am_g : g \in [G]\}$ is bound to increase.
Thus, there exists an upper bound on $G$ for which decoding succeeds with high probability. 
An exact characterization of such a bound remains an open research question.
Still, these notions provide guiding principles on the selection of $G$.
\end{remark}
\begin{remark}
When $G = 1$, every active user belongs to the same bin.
Thus, under the URA common task framework, $K$ is known and there is no need to perform occupancy estimation. 
In this situation, active users would refrain from sending a bin identification sequence and would instead employ $100\,\%$ of their power for transmitting their payloads. 
Under these modifications, the $G=1$ coded demixing scheme reduces to the CCS-AMP scheme of Amalladinne et al.\ in \cite{amalladinne2020unsourced}.
\end{remark}

\subsection{Single-Class Simulation Results}

We now characterize the performance of a single-class URA system that employs coded demixing and stochastic binning.
We consider competing implementations with the parameters summarized in Table~\ref{table:cdresults}.
The power budget $P$ is obtained via the relationship $\frac{E_{\mathrm{b}}}{N_0} = \frac{nP}{2w}$ and, as discussed in Remark~\ref{remark:hadamardmatrices}, the sensing matrices $\{\Am_g : g \in [G]\}$ are generated by randomly sampling the rows of a $2^v \times 2^v$ Hadamard matrix, excluding the row of all ones. 

\begin{table}[t]
    \centering
    \caption{Summary of simulation parameters used to generate Fig.~\ref{fig:PupevsBins}}
    \label{table:cdresults}
    \begin{tabular}{||c|c|c|c|c|c||}
        \hline
        Scheme & $G$ & $K$ & $L$ & $w$ & $n$ \\
        \hline\hline
         CCS-AMP & 1 & 25-175 & 16 & 128 & 38400  \\
         \hline
         CCS-AMP & 1 & 200-275 & 18 & 128 & 38400 \\ 
         \hline
         Coded Demixing & 2 & 25-275 & 16 & 128 & 38400 \\ 
         \hline
         Coded Demixing & 8 & 25-275 & 16 & 128 & 38400 \\
         \hline
    \end{tabular} 
\end{table}

\begin{figure}[ht]
    \centering
    \begin{tikzpicture}

\definecolor{vamsired}{rgb}{0.63529,0.07843,0.18431} % red
\definecolor{vamsiblue}{rgb}{0.00000,0.44706,0.74118} % blue
\definecolor{vamsigreen}{rgb}{0.00000,0.49804,0.00000} % dark green
\definecolor{vamsiorange}{rgb}{0.87059,0.49020,0.00000} % orange
\definecolor{mycolor5}{rgb}{0.00000,0.44700,0.74100} %
\definecolor{mycolor6}{rgb}{0.74902,0.00000,0.74902} %

\begin{axis}[%
font=\small,
width=7cm,
height=5.5cm,
scale only axis,
every outer x axis line/.append style={white!15!black},
every x tick label/.append style={font=\color{white!15!black}},
xmin=25,
xmax=275,
xtick = {0,50,100,...,275},
xlabel={Number of active users $K$},
xmajorgrids,
every outer y axis line/.append style={white!15!black},
every y tick label/.append style={font=\color{white!15!black}},
ymin=1,
ymax=5,
ytick = {1,...,5},
ylabel={Required $E_{\mathrm{b}}/N_0$ (dB)},
ymajorgrids,
legend style={at={(0,1)},anchor=north west, draw=black,fill=white,legend cell align=left}
]

% \addplot [color=black,dotted,line width=1.5pt]
%   table[row sep=crcr]{
%  25	0.25\\
% 50	0.3\\
% 75	0.35\\
% 100	0.4\\
% 125	0.45\\
% 150	0.5\\
% 175	0.55\\
% 200	0.6\\
% 225	0.95\\
% 250	1.25\\
% 275	1.55\\
% % 300	1.8\\
% };
% \addlegendentry{Random Coding \cite{polyanskiy2017perspective}};

\addplot [color=vamsigreen,solid,line width=2.0pt,mark size=1.4pt,mark=o,mark options={solid}]
  table[row sep=crcr]{10 1.7 \\
25 1.85 \\
50 2.08 \\
75 2.31 \\
100 2.38 \\
125 2.65\\
150 2.99 \\
175 3.12 \\
200 3.57\\
% 225 3.6 \\
% 250 3.82 \\
% 275 4.31 \\
% 300 4.89 \\
};
\addlegendentry{CCS-AMP \cite{amalladinne2020unsourced}};

% JRE NOTE: this corresponds to the original curve. These results are outdated and sub-optimal
% \addplot [color=red, dashed,line width=2.0pt,mark size=1.4pt,mark=o,mark options={solid}]
% table[row sep=crcr]{
%   25 1.86 \\
%   50 1.92 \\
%   75 1.95 \\
%   100 2.05 \\
%   125 2.08 \\
%   150 2.24 \\
%   175 2.4 \\
%   200 2.6 \\
%   225 2.79 \\
%   250 3.15 \\
%   275 3.75 \\
% };
% \addlegendentry{Coded Demixing, $G = 2$};

% JRE Note: this corresponds to best-known results
\addplot [color=red, solid,line width=2.0pt,mark size=1.4pt,mark=o,mark options={solid}]
table[row sep=crcr]{
  25 1.60 \\
  50 1.64 \\
  75 1.72 \\
  100 1.79 \\
  125 1.97 \\
  150 2.15 \\
  175 2.32 \\
  200 2.53 \\
  225 2.79 \\
  250 3.15 \\
  275 3.75 \\
};
\addlegendentry{Coded Demixing, $G = 2$};

% JRE NOTE: this corresponds to the original curve. These results are outdated and sub-optimal
% \addplot [color=vamsiblue, dashed,line width=2.0pt,mark size=1.4pt,mark=o,mark options={solid}]
% table[row sep=crcr]{
%   25 1.71 \\
%   50 1.80 \\
%   75 1.83 \\
%   100 1.91 \\
%   125 1.98 \\
%   150 2.02 \\
%   175 2.13 \\
%   200 2.43 \\
%   225 3.15 \\
%   250 3.89 \\
%   275 4.93 \\
% };
% \addlegendentry{Coded Demixing, $G = 8$};

% JRE Note: this corresponds to best-known results
\addplot [color=vamsiblue, solid,line width=2.0pt,mark size=1.4pt,mark=o,mark options={solid}]
table[row sep=crcr]{
  25 1.64 \\
  50 1.691 \\
  75 1.72 \\
  100 1.77 \\
  125 1.81 \\
  150 1.89 \\
  175 2.00 \\
  200 2.41 \\
  225 3.15 \\
  250 3.89 \\
  275 4.93 \\
};
\addlegendentry{Coded Demixing, $G = 8$};

\addplot [color=vamsigreen,dashed,line width=2.0pt,mark size=1.4pt,mark=o,mark options={solid}]
  table[row sep=crcr]{
% 10 1.7 \\
% 25 1.85 \\
% 50 2.08 \\
% 75 2.31 \\
% 100 2.38 \\
% 125 2.65\\
% 150 2.99 \\
% 175 3.12 \\
200 3.57\\
225 3.6 \\
250 3.82 \\
275 4.31 \\
% 300 4.89 \\
};

\end{axis}
\end{tikzpicture}
    \caption{Required $E_{\mathrm{b}}/N_0$ for PUPE = $0.05$ vs number of active users $K$ for $G = 1, 2, 8$. 
    Under coded demixing, adding bins improves the performance of the URA system, up to a point (see Remark~\ref{remark:binlimits}).}
    \label{fig:PupevsBins}
\end{figure}
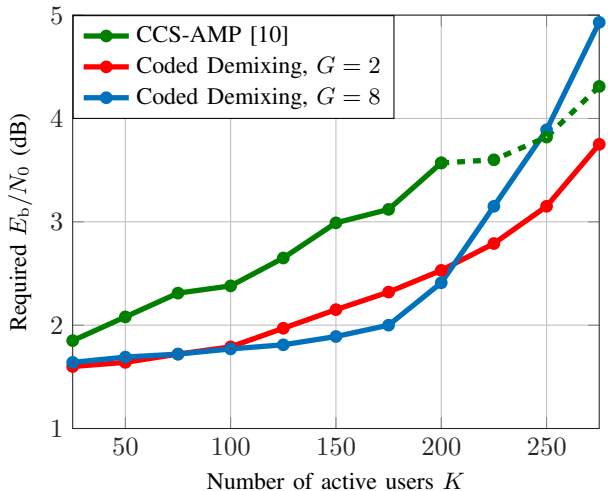

Fig.~\ref{fig:PupevsBins} compares the required $E_{\mathrm{b}}/N_0$ to obtain a PUPE of $0.05$ as a function of $K$ for $G = 1, 2, 8$.
The curve for $G = 1$ corresponds to the original CCS-AMP scheme presented in \cite{amalladinne2020unsourced}. 
Fig.~\ref{fig:PupevsBins} empirically reveals that the $G = 2$ curve uniformly outperforms CCS-AMP for all $K$, and that the $G = 8$ curve outperforms both CCS-AMP and $G = 2$ for moderate $K$ (i.e., $75 \leq K \leq 200)$.
However, when $G = 8$ and $K$ is large, the performance of the $G = 8$ curve decays sharply. 

There are some nuances to Fig.~\ref{fig:PupevsBins} that need to be considered when interpreting the results. 
Recall that, as part of the decoding process, BP is run on the outer factor graph to disentangle the $K$ users present on that graph. 
Empirical results evidence an upper limit on $K$ for which the decoding process performs arbitrarily well as $E_{\mathrm{b}}/N_0 \rightarrow \infty$; in our case, this upper limit is around $200$.
Thus, when generating the CCS-AMP curve, the outer graph has to be extended to $18$ sections for $K \geq 200$ in order to obtain a PUPE of $0.05$.
Furthermore, the extended graph is characterized by a crossover SNR when compared to the original graph.
For the region of $K$ and $E_{\mathrm{b}}/N_0$ considered in Fig.~\ref{fig:PupevsBins}, the extended graph did not improve the performance of the $G = 2$ or $G = 8$ bin curves and, therefore, it was not used for $K \geq 200$.
The CCS-AMP curve is displayed, partially, as a dotted line on this figure to highlight the region where the extended outer graph is being used.
Altogether, the parameter space is vast and the performance of coded demixing appears to be sensitive to the choice of parameters. 
The optimal choice of parameters, including the optimal design of the outer graph, remains an open research problem. 

\begin{figure}[ht!]
    \centering
    \begin{tikzpicture}

\definecolor{vamsired}{rgb}{0.63529,0.07843,0.18431} % red
\definecolor{vamsiblue}{rgb}{0.00000,0.44706,0.74118} % blue
\definecolor{vamsigreen}{rgb}{0.00000,0.49804,0.00000} % dark green
\definecolor{vamsiorange}{rgb}{0.87059,0.49020,0.00000} % orange
\definecolor{mycolor5}{rgb}{0.00000,0.44700,0.74100} %
\definecolor{mycolor6}{rgb}{0.74902,0.00000,0.74902} %

\begin{axis}[%
font=\small,
width=7cm,
height=5.5cm,
scale only axis,
every outer x axis line/.append style={white!15!black},
every x tick label/.append style={font=\color{white!15!black}},
xmin=25,
xmax=275,
xtick = {0,50,100,...,275},
xlabel={Number of active users $K$},
xmajorgrids,
every outer y axis line/.append style={white!15!black},
every y tick label/.append style={font=\color{white!15!black}},
ymin=0,
ymax=6,
ytick = {0,...,6},
ylabel={Required $E_{\mathrm{b}}/N_0$ (dB)},
ymajorgrids,
legend style={at={(0,1)},anchor=north west, draw=black,fill=white,legend cell align=left}
]

\addplot [color=black,dotted,line width=1.5pt]
  table[row sep=crcr]{
 25	0.25\\
50	0.3\\
75	0.35\\
100	0.4\\
125	0.45\\
150	0.5\\
175	0.55\\
200	0.6\\
225	0.95\\
250	1.25\\
275	1.55\\
% 300	1.8\\
};
\addlegendentry{Random Coding \cite{polyanskiy2017perspective}};

\addplot [color=vamsigreen,solid,line width=2.0pt, mark=o,mark options={solid}]
  table[row sep=crcr]{
10 1.7 \\
25 1.85 \\
50 2.08 \\
75 2.31 \\
100 2.38 \\
125 2.65\\
150 2.99 \\
175 3.12 \\
200 3.57\\
225 3.6 \\
250 3.82 \\
275 4.31 \\
% 300 4.89 \\
};
\addlegendentry{CCS-AMP \cite{amalladinne2020unsourced}};

% \addplot [color=mycolor5,solid,line width=2.0pt,mark size=1.4pt,mark=o,mark options={solid}]
%   table[row sep=crcr]{
%   25  0.1\\
%   50 0.1\\
%   75 0.3
%   100 0.4\\
%   125 0.55\\
%   150 1\\
%   175 1.3\\
% 200 1.65\\
% 225 2.35\\
% 250 2.85\\
% 275 3.43\\
% };
% \addlegendentry{LDPC + SC \cite{pradhan2021ldpc}};

\addplot [color=mycolor5,solid,line width=2.0pt,mark size=1.4pt,mark=o,mark options={solid}]
  table[row sep=crcr]{
  25 0.29 \\
  50 0.32 \\
  75 0.43 \\
  100 0.58 \\
  125 0.85 \\
  150 1.25 \\
  175 1.70 \\
  200 2.25 \\
  225 2.80 \\
  250 3.20 \\
  275 3.85 \\
};
\addlegendentry{SKP Coding \cite{han2021sparse}};

\addplot [color=mycolor6,solid,line width=2.0pt,mark size=1.4pt,mark=o,mark options={solid}]
  table[row sep=crcr]{
  25  2\\
50	2.1\\
75	2.2\\
100	2.41\\
125	2.57\\
150	2.81\\
175	3\\
200 3.4\\
225 3.88\\
250 4.36\\
275 4.87\\
300 5.35\\
};
\addlegendentry{Sparse IDMA \cite{pradhan2020sparse}};

\addplot [color=vamsiorange,solid,line width=2.0pt,mark size=1.4pt,mark=o,mark options={solid}]
  table[row sep=crcr]{
  2     0.3\\
  10    0.5\\
 25	0.55\\
50	0.6\\
75 0.7\\
100 0.75\\
125 1.15\\
150 1.5\\
175 2\\
200 2.7\\
225 3.5\\
250 4.3\\
};
\addlegendentry{Polar+Spreading \cite{pradhan2019polar}};

\addplot [color=red, solid,line width=2.0pt,mark size=1.4pt,mark=o,mark options={solid}]
table[row sep=crcr]{
  25 1.60 \\
  50 1.64 \\
  75 1.72 \\
  100 1.79 \\
  125 1.97 \\
  150 2.15 \\
  175 2.32 \\
  200 2.53 \\
  225 2.79 \\
  250 3.15 \\
  275 3.75 \\
};
\addlegendentry{Coded Demixing, $G = 2$};

\end{axis}
\end{tikzpicture}
    \caption{This plot shows required $E_{\mathrm{b}}/N_0$ for PUPE = $0.05$ vs number of active users $K$. The performance of coded demixing with $G = 2$ is compared to other URA schemes.}
    \label{fig:PupevsScheme}
\end{figure}
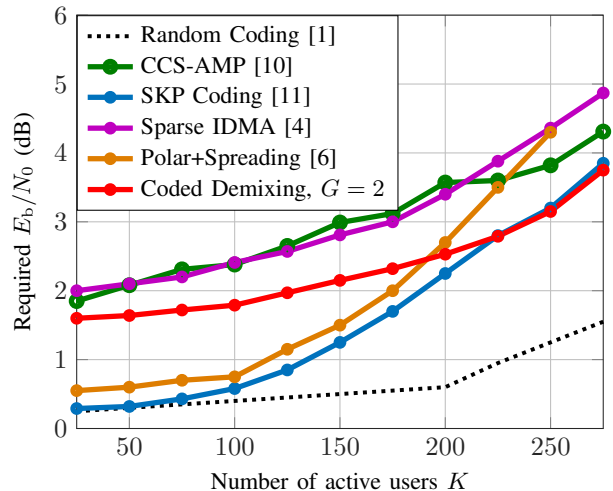

Fig.~\ref{fig:PupevsScheme} compares the required $E_{\mathrm{b}}/N_0$ to obtain a PUPE of $0.05$ as a function of $K$ for coded demixing to Polyanskiy's random coding benchmark~\cite{polyanskiy2017perspective}, CCS-AMP \cite{amalladinne2020unsourced}, Sparse Kronecker-Product coding \cite{han2021sparse}, Sparse IDMA \cite{pradhan2020sparse}, and polar coding with spreading \cite{pradhan2019polar}. 
When $G = 2$ and $K \geq 225$, coded demixing outperforms all referenced schemes in terms of required $E_{\mathrm{b}}/N_0$. 

\subsection{Performance with $K$ Unknown at Receiver}

Throughout this article, we have assumed that the total number of active devices $K$ is known at the receiver.
This assumption is in line with existing URA literature and allows for a fair comparison with existing schemes.
However, in many scenarios of practical interest, this assumption is not properly justified; therefore, we briefly consider the case when $K$ is unknown a priori and must be estimated. 
Interestingly, under minor modifications, coded demixing still performs well in this alternate scenario. 
This should not be too surprising in view of recently published results on URA without side information~\cite{ngo2021massive}, a study which concludes that the lack of knowledge in the number of active devices should entail only a small penalty in power efficiency.

When $K$ is not known at the receiver, the total number of active devices can be inferred from the bin identification sequences.
Because our proposed LMMSE estimator relies on $K$ as known parameter, a different estimator should be employed. 
To obtain $\{\hat{K}_g : g \in [G]\}$, the receiver may simply divide the received bin identification sequence by the users' transmit power and then round the result to the nearest integer. 
The estimated total number of active users $\hat{K}$ is then given by $\hat{K} = \sum_{g \in [G]} \hat{K}_g$. 
The rest of the coded demixing algorithm proceeds exactly as has been described in this article, albeit using $\hat{K}$ instead of $K$.

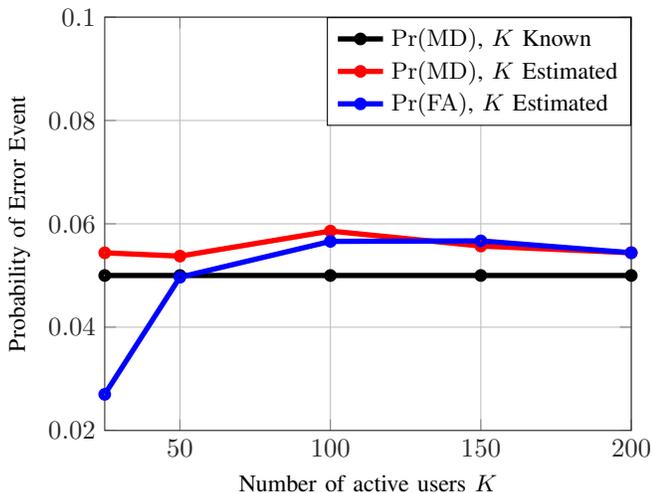
\begin{figure}
    \centering
    \begin{tikzpicture}

\definecolor{vamsired}{rgb}{0.63529,0.07843,0.18431} % red
\definecolor{vamsiblue}{rgb}{0.00000,0.44706,0.74118} % blue
\definecolor{vamsigreen}{rgb}{0.00000,0.49804,0.00000} % dark green
\definecolor{vamsiorange}{rgb}{0.87059,0.49020,0.00000} % orange
\definecolor{mycolor5}{rgb}{0.00000,0.44700,0.74100} %
\definecolor{mycolor6}{rgb}{0.74902,0.00000,0.74902} %

\begin{axis}[%
font=\small,
width=7cm,
height=5.5cm,
scale only axis,
every outer x axis line/.append style={white!15!black},
every x tick label/.append style={font=\color{white!15!black}},
scaled ticks=false,
tick label style={/pgf/number format/fixed},
xmin=25,
xmax=200,
xtick = {0,50,100,...,275},
xlabel={Number of active users $K$},
xmajorgrids,
every outer y axis line/.append style={white!15!black},
every y tick label/.append style={font=\color{white!15!black}},
ymin=0.02,
ymax=0.10,
ytick = {0.02, 0.04, 0.06, 0.08, 0.10},
ylabel={Probability of Error Event},
ymajorgrids,
legend style={at={(1,1)},anchor=north east, draw=black,fill=white,legend cell align=left}
]

\addplot [color=black, solid,line width=2.0pt,mark size=1.4pt,mark=o,mark options={solid}]
table[row sep=crcr]{
  25 0.05 \\
  50 0.05 \\
  100 0.05 \\
  150 0.05 \\
  200 0.05 \\
  250 0.05 \\
};
\addlegendentry{$\Pr(\mathrm{MD})$, $K$ Known};

\addplot [color=red, solid,line width=2.0pt,mark size=1.4pt,mark=o,mark options={solid}]
table[row sep=crcr]{
  25 0.05439 \\
  50 0.05375 \\
  100 0.0586 \\
  150 0.0557 \\
  200 0.0544 \\
  250 0.0424 \\
};
\addlegendentry{$\Pr(\mathrm{MD})$, $K$ Estimated};

\addplot [color=blue, solid,line width=2.0pt,mark size=1.4pt,mark=o,mark options={solid}]
table[row sep=crcr]{
  25 0.027 \\
  50 0.04967 \\
  100 0.0566 \\
  150 0.0567 \\
  200 0.0544 \\
  250 0.0417 \\
};
\addlegendentry{$\Pr(\mathrm{FA})$, $K$ Estimated};

\end{axis}
\end{tikzpicture}
    \caption{This figure showcases the $\Pr\left(\mathrm{MD}\right)$ and $\Pr\left(\mathrm{FA}\right)$ of coded demixing when $G = 2$ for the case when $K$ is unknown at the receiver. Transmit powers are selected to achieve $\Pr\left(\mathrm{MD}\right) = 0.05$ when $K$ is known, serving as a performance benchmark.
    Clearly, not knowing $K$ at the receiver incurs only a small performance penalty. }
    \label{fig:pe_unknown_k}
\end{figure}

When $K$ is unknown, we consider both the probability of missed detection $\Pr\left(\mathrm{MD}\right)$, defined as the probability of a user's message not appearing in $\hat{\mathcal{W}}\left(\yv\right)$, and the probability of false alarm $\Pr\left(\mathrm{FA}\right)$, defined as the probability of a codeword appearing in $\hat{\mathcal{W}}\left(\yv\right)$ that does not correspond to a transmitted message. 
In contrast, when $K$ is known and used as a constraint on the size of the output list, $\Pr\left(\mathrm{MD}\right) = P_e$ and $\Pr\left(\mathrm{FA}\right) \leq P_e$. 
Fig.~\ref{fig:pe_unknown_k} shows the performance of coded demixing when $K$ is unknown at the receiver for $G = 2$ at the $E_{\mathrm{b}}/N_0$ values identified in Fig.~\ref{fig:PupevsBins}. 
From this figure, it is clear that not knowing $K$ incurs only a small performance penalty in terms of $\Pr\left(\mathrm{MD}\right)$ and $\Pr\left(\mathrm{FA}\right)$. 
We point out that, despite the fact that the receiver does not have exact knowledge of $K$, the devices themselves implicitly assume that $K \leq 275$.
If $K$ were to grow much larger, the code rate would eventually have to decrease as it did for the CCS-AMP curve in Fig.~\ref{fig:PupevsBins}.
This broad assumption about user density is akin to those employed in cellular systems.

\section{Conclusion}
\label{section:conclusion}

%  In this article, a novel URA scheme is proposed that incorporates ideas from the field of \textcolor{black}{convex demixing} into coded compressed sensing (CCS).
In this article, a novel signal processing tool called coded demixing is presented that extends existing notions of convex demixing and coded compressed sensing (CCS) to enable the joint recovery of very high dimensional signals that are sparse with respect to separate bases. 
A pragmatic framework featuring an approximate message passing (AMP) based decoder is developed that enables all sent signals to be disentangled and for individual sparsity patterns to be recovered accurately with low-complexity. 

This novel framework may be utilized in a heterogeneous multi-class URA setting where each class features distinct power, coding, and data requirements.  
In such a network, coded demixing is shown to significantly outperform traditional techniques such as treating interference as noise and successive interference cancellation. 
Somewhat surprisingly, coded demixing with stochastic binning may also be used to enhance single-class URA networks. 
This approach significantly improves PUPE performance because the number of users present on a single bipartite factor graph is dramatically reduced, which in turns boosts the performance of the BP decoder. 
It is shown that increasing the number of bins of users improves performance, up to a point. 
When the number of bins is too large, the operating point of the system may fall out of AMP's convergence region and/or the cross-coherence of the sensing matrices may grow too large for reliable demixing. 

Coded demixing has a promising future within URA. 
Open research challenges include precisely characterizing the upper bound on the number of bins that may be employed within a single-class system and finding an optimal scheme for the heterogenous multi-class URA setting. 
Furthermore, incorporating fading into the proposed framework as well as extending the framework to accommodate multiple input multiple output (MIMO) scenarios remain open research directions.

\appendices

\section{Derivation of Onsager Term}
\label{appendix:onsagerderivation}

In this section, we derive the form of the Onsager term for the coded demixing algorithm of Section~\ref{section:CodedDemixing}.
Recall that the dynamic denoiser used within AMP is defined as
\begin{equation} \label{eq:dynamicdenoiser_appendix}
     \etav^{(t)}\left(\rv^{(t)}\right)
     = \begin{bmatrix} \etav_1^{(t)}\left(\rv_1^{(t)}\right) \\
    \vdots \\ \etav_G^{(t)}\left(\rv_G^{(t)}\right) \end{bmatrix},
\end{equation}
where the group denoisers are of the form
\begin{equation}
    \etav^{(t)}_{g}(\rv_g) = \hat{\sv}_{g}(1, \rv_g, \tau_t) \cdots \hat{\sv}_{g}(L_g, \rv_g, \tau_t).
\end{equation}
Elements within the denoiser for group~$g$ are given by
\begin{equation}
    \label{eq:jointpme_appendix}
    \hat{\sv}_{g}(\ell, \rv_g(\ell), \tau_t) = \left( \hat{s}_{g}(\qv_g(\ell, k), \rv_g(\ell, k), \tau_t) : k \in [2^{v_g}] \right)
\end{equation}
and $\hat{s}_{g} \left( q, r, \tau_t \right)$ is the posterior mean estimate (PME) equal to
\begin{equation} \label{eq:pme_appendix}
    \begin{split}
        \hat{s}_{g} \left( q, r, \tau_t \right) &=     \\
        &\frac{q \exp \left( - \frac{ \left( r - d_{g} \right)^2}{2 \tau_t^2} \right)}
        { q \exp \left( - \frac{ \left( r -  d_{g} \right)^2}{2 \tau_t^2} \right)
        + (1-q) \exp \left( -\frac{r^2}{2 \tau_t^2} \right)}, \\
    \end{split}
\end{equation}
We emphasize that this denoiser is utilized twice within the AMP iterate: once during the state update to promote sparsity, and once during the residual computation as part of the Onsager correction term.
Below, we formally derive the needed expression for the Onsager term,
\begin{equation}
    \label{eq:onsagercorrectionterm_appendix}
    \frac{\zv}{n}\mathrm{div}\Dm\etav^{(t)} = \frac{\zv}{n \tau_t^2}\left(
    \| \Dm^2\etav^{(t)}(\rv) \|_1 - \| \Dm\etav^{(t)}(\rv) \|_2^2 \right) .
\end{equation}
Before doing so, we first consider the partial derivative of \eqref{eq:pme_appendix}, which is required in the computation of the Onsager term. 

\begin{lemma} [Adapted from \cite{amalladinne2020unsourced}, Lemma 7]
\label{lemma:partialderivative_pme_appendix}
If the number of BP iterations on the outer factor graph is strictly less than the length of its shortest cycle, then the partial derivative of $\hat{s}_{g}(\qv(\ell, k), \rv(\ell, k), \tau_{t})$ with respect to $\rv(\ell, k)$ is equal to
\begin{equation} \label{equation:partialderivative_pme_appendix}
    % \frac{\partial \hat{s}_{\ell} \left( \qv(\ell, k), \rv(\ell, k), \tau_{\ell} \right)}{\partial \rv \left( \ell, k \right)}
    % = \
    \frac{d_{g}}{\tau_t^2} \hat{s}_{g}\left( \qv(\ell, k), \rv(\ell, k), \tau_{t} \right)  
    \left( 1 - \hat{s}_{g} \left( \qv(\ell, k), \rv(\ell, k), \tau_{t} \right) \right) .
\end{equation}
\end{lemma}
\begin{IEEEproof}
For simplicity of notation, we use $q = \qv(\ell, k)$, $r = \rv(\ell, k)$, and $\tau = \tau_{t}$.
Then, we can write
\begin{equation*}
    \begin{split}
        \hat{s}_{g}(q, r, \tau)
        & = \frac{q \exp \left( - \frac{ \left( r - d_{g} \right)^2}{2 \tau^2} \right)}{ q \exp \left( - \frac{ \left( r -  d_{g} \right)^2}{2 \tau^2} \right) + (1-q) \exp \left( -\frac{r^2}{2 \tau^2} \right)} \\
        & = \frac{q}{q + (1 - q) \exp \left( \frac{d_{g}^2 - 2rd_{g}}{2\tau^2} \right)} .
    \end{split}
\end{equation*}
Because the number of BP iterations is less than the length of the shortest cycle, ${\partial q} / {\partial r}  = 0$.  Thus, using the chain rule of differentiation, we obtain
\begin{equation*}
    \begin{split}
        \frac{\partial \hat{s}_{g}(q, r, \tau)}{\partial r}
        &= \frac{d_{g}}{\tau^2} 
        \frac{q(1-q)\exp \left( \frac{d_{g}^2 - 2 r d_{g}}{2 \tau^2} \right)}{\left( q + (1 - q) \exp \left( \frac{ d_{g}^2 - 2 r d_{g}}{2 \tau^2} \right) \right)^2} \\
        &= \frac{d_{g}}{\tau^2}\hat{s}_{g}(q, r, \tau)(1 - \hat{s}_{g}(q, r, \tau)) .
    \end{split}
\end{equation*}
In this context, $q$ acts as a prior or belief inherited from neighboring sections.
When the number of BP iterations on the LDPC factor graph is strictly less than the length of its shortest cycle, the computation of $\qv(\ell, k)$ does not depend on the value of $\rv(\ell, k)$.
This assumption is crucial in obtaining the expression above, else the partial derivative of $\qv(\ell, k)$ with respect to $\rv(\ell, k)$ would have to be considered as well.
\end{IEEEproof}

Having established Lemma~\ref{lemma:partialderivative_pme_appendix}, we are ready to compute the divergence of the denoiser.  

\begin{proposition}
    \label{proposition:denoiserdivergence_appendix}
    The divergence of $\Dm\etav^{(t)}(\rv)$ with respect to $\rv$ is given by 
    \begin{equation}
        \mathrm{div} \Dm\etav^{(t)}(\rv) = \frac{1}{\tau_t^2}\left( \| \Dm^2\etav^{(t)}(\rv) \|_1 - \| \Dm\etav^{(t)}(\rv) \|_2^2 \right) ,
    \end{equation}
    provided that the number of BP iterations on the outer factor graph is strictly less than the length of the shortest cycle.
\end{proposition}
\begin{IEEEproof}
The divergence of $\Dm\etav^{(t)}(\rv)$ may be expressed as
\begin{equation*}
    \begin{split}
        \mathrm{div}\Dm\etav^{(t)}(\rv) &=
        \sum_{g = 1}^{G} d_{g} \mathrm{div} \hspace{0.5mm} \etav^{(t)}_{g}(\rv_{g}) \\
        &=\sum_{g = 1}^{G} d_{g} \mathrm{div} \left( \hat{\sv}_g(1, \rv_g, \tau_t)\cdots\hat{\sv}_g(L_g, \rv_g, \tau_t) \right) \\
        % &= \sum_{g = 1}^{G} d_{g} \sum_{\ell = 1}^{L} \mathrm{div} \hspace{0.5mm} \hat{\sv}_{\ell}(\rv_g, \tau_t) \\
        &= \sum_{g = 1}^{G} d_{g} \sum_{\ell = 1}^{L_g} \sum_{k = 1}^{2^v} \frac{\partial \hat{\sv}_{g, k}(\ell, \rv_g, \tau_t)}{\partial \rv_g(\ell, k)} .
    \end{split}
\end{equation*}
Under the assumption that the number of BP iterations on the outer factor graph is strictly less than the length of the shortest cycle and using the result of Lemma \ref{lemma:partialderivative_pme_appendix}, we get
    \begin{equation*}
        \begin{split}
            &\mathrm{div}\Dm\etav^{(t)}(\rv) \\
            &= \sum_{g = 1}^{G} \sum_{\ell = 1}^{L_g} \sum_{k = 1}^{2^v} \frac{d_{g}^2}{\tau_t^2} \hat{\sv}_{g, k}\left(\rv_{g}(\ell), \tau_t \right) \left( 1 - \hat{\sv}_{g, k} \left(\rv_{g}(\ell), \tau_t \right) \right) .
        \end{split}
    \end{equation*}
The above expression may be represented as the sum of $G$ inner products,
    \begin{equation*}
        \begin{split}
            \mathrm{div}\Dm\etav^{(t)}(\rv) &= \sum_{g = 1}^{G}\frac{1}{\tau_t^2} \langle d_{g} \etav^{(t)}_{g}(\rv_{g}), d_{g}\onev - d_{g}\etav^{(t)}_{g}(\rv_{g})\rangle \\
            &= \sum_{g = 1}^{G} \frac{1}{\tau_t^2}\left( \|d_g^2 \etav_g^{(t)}(\rv_g)\|_{1} -\|d_g \etav_{g}^{(t)}(\rv_g)\|_2^2 \right) .
        \end{split}
    \end{equation*}
    Leveraging the short-hand forms
    \begin{align*}
    \left\| \Dm^2 \etav^{(t)}(\rv) \right\|_{1} = \sum_{g = 1}^{G} \left\| d_g^2\etav_{g}^{(t)}(\rv_g) \right\|_{1} \\
    \left\| \Dm \etav^{(t)}(\rv) \right\|_2^2 = \sum_{g = 1}^{G} \left\| d_g \etav_g^{(t)}(\rv_g) \right\|_2^2 ,
    \end{align*}
    the divergence of $\Dm\etav^{(t)}(\rv)$ simplifies to
    \begin{equation*}
        \mathrm{div}\Dm\etav^{(t)}(\rv) = \frac{1}{\tau_t^2}\left( \|\Dm^2 \etav^{(t)}(\rv)\|_1 - \|\Dm \etav^{(t)}(\rv)\|_2^2 \right) .
    \end{equation*}
This completes the proof.
\end{IEEEproof}

The full Onsager term in \eqref{eq:onsagercorrectionterm_appendix} follows directly from the results of Proposition~\ref{proposition:denoiserdivergence_appendix}. 

\bibliographystyle{IEEEbib}
\bibliography{tsp-28432-2021}

\end{document}